\documentclass[a4paper,11pt]{article}
\usepackage{aaskaiid}
\usepackage{aas_macros}%
\usepackage[utf8]{inputenc}
\hypersetup{
    colorlinks=true,
    linkcolor=red,
    filecolor=magenta,      
    urlcolor=magenta,
    citecolor=blue,
}
\usepackage{multirow}
\usepackage{orcidlink}
\DeclareUnicodeCharacter{2212}{-} 
\usepackage{tabularx} 
\usepackage{adjustbox} 
\usepackage{pdflscape} 
\usepackage{graphicx} 
\usepackage{lscape} 
\usepackage{longtable} 
\usepackage{booktabs} 
\usepackage{caption} 
\usepackage{adjustbox}
\usepackage{threeparttable}

\title{Solar Radio Bursts in the metric to kilometric range}
\ShortTitle{Solar Radio Bursts}

\author[1]{Anshu Kumari\orcidlink{0000-0001-5742-9033}}
\ShortName{Kumari et al.} 
\author[2]{Mugundhan V.\orcidlink{0000-0003-3630-3065}}
\author[3,4]{Diana E. Morosan\orcidlink{0000-0002-8416-1375}} 
\author[5,6]{Jasmina Magdalenić\orcidlink{0000-0003-1169-3722}}
\author[5,6]{Ketaki Deshpande\orcidlink{0000-0001-6861-6328}}
\author[7,8]{Peijin Zhang\orcidlink{0000-0001-6855-5799}}
\author[1,9]{Divya Paliwal\orcidlink{0009-0001-7689-0084}}
\author[10]{Pietro Zucca\orcidlink{0000-0002-6760-797X}}
\author[11]{Puja Majee \orcidlink{0000-0002-2711-2366}}

\affiliation[1]{Udaipur Solar Observatory, Physical Research Laboratory, Dewali, Badi Road, Udaipur - 313001, Rajasthan, India}
\emailAdd{anshu@prl.res.in}
\affiliation[2]{Dept. of Space Planetary Astronomical Sciences and Engineering (SPASE), IITK, Kanpur 208016, India}
\affiliation[3]{Department of Physics and Astronomy, University of Turku, 20014, Turku, Finland}
\affiliation[4]{Turku Collegium for Science, Medicine and Technology, University of Turku, 20014, Turku, Finland}
\affiliation[5]{Center for mathematical Plasma Astrophysics (CmPA), KU Leuven, Celestijnenlaan 200B, 3001 Leuven, Belgium}
\affiliation[6]{Solar-Terrestrial Centre of Excellence (STCE), Royal Observatory of Belgium, Avenue Circulaire 3, 1180 Uccle, Belgium}
\affiliation[7]{Center for Solar-Terrestrial Research, New Jersey Institute of Technology, Newark, NJ 07102, USA}
\affiliation[8]{Cooperative Programs for the Advancement of Earth System Science, University Corporation for Atmospheric Research, Boulder, CO, USA}
\affiliation[9]{Indian Institute of Technology, Gandhinagar, Gujarat-382355, India}
\affiliation[10]{ASTRON, The Netherlands Institute for Radio Astronomy, Oude Hoogeveensedijk 4, 7991 PD Dwingeloo, The Netherlands}
\affiliation[11]{National Centre for Radio Astrophysics, Tata Institute of Fundamental Research, S. P. Pune University Campus, Pune 411007, India}

\abstract{Solar radio bursts (SRBs) are intense emissions observed in radio wavelengths most frequently during solar transients, such as coronal mass ejections (CMEs) and flares. SRBs are direct signatures of accelerated electrons in the solar atmosphere. These solar transients have a direct impact on the near-Earth atmosphere. SRBs serve as key diagnostic tools for plasma processes, particle accelerations,  magnetic field dynamics in the solar corona and the heliosphere, which are the root cause of these solar transients. There are several key science question which solar radio observations can answer, such as: When $\&$ where is the bulk of the energy released in flares?, what are the physical properties of the energy release site?, what are the properties of heated plasma $\&$ accelerated particles?, how does the transport of heated plasma $\&$ accelerated particles?, what bearing do flares have on the question of coronal heating? The Square Kilometre Array (SKA), with its unprecedented sensitivity, temporal, spectral, and spatial resolution, as well as dynamic range, is expected to provide an enhanced understanding of the physics behind solar transients with unprecedented detail. }

\begin{document}
\maketitle


\section{Introduction} 

Solar radio emission was historically divided into three main components: emission originating from the quiet Sun, emission associated with bright active regions, and radio emission associated with solar eruptive phenomena. Herein, we focus on the metric and decametric solar radio bursts (< 500~MHz), which is generally of a significantly higher intensity than the quiet Sun emission. Solar radio bursts are associated with either the emission of active regions or eruptive phenomena, such as solar flares and coronal mass ejections (CMEs). Radio bursts are generally classified based on the spectral appearance of the bursts, and the wavelength regimes of their appearance \citep{wild1950, SuzukiDulk85, nelson1985, McLean1985}, which is closely related to the emission mechanisms of the radio bursts \citep{1961ApJ...133..983N, 1980SSRv...26....3M,1985ARA&A..23..169D}. Considering the morphological characteristics, five main categories of solar radio bursts can be distinguished: Type I, Type II, Type III, Type IV, and Type V. A detailed description is provided in Section~\ref{active_sun_obs}. 

Solar radio bursts are observed across a wide range of wavelengths, from millimetres to kilometres, corresponding to frequencies from the GHz to the kHz range. Observations at short wavelengths correspond to distances close to the solar surface, while those at long wavelengths correspond to the large distances from the Sun. This means that radio bursts map the processes occurring at nearly all levels of the solar corona, extending as far away from the Sun as interplanetary space. Such a broad coverage makes solar radio bursts a powerful diagnostic tool for mapping the coronal plasma properties \cite[see e.g.][]{vr01,vr02,vr03,Kumari2019PhD, deshpande25}, or tracking the propagation of their exciters \cite[see e.g.][]{Magdalenic2008, magdalenic14, kumari2017a, Zhang2019type3durationPosition,kumari2017b, kumari2017c,jebaraj2020,morosan2022,Pietro2025Multilane}.
At microwave frequencies, solar radio bursts are typically dominated by gyrosynchrotron emission from mildly relativistic electrons in flaring loops. These phenomena provide complementary diagnostics of magnetic fields and energetic particles but are not treated in detail here.

The Square Kilometre Array (SKA) will transform solar and heliospheric radio science by offering unprecedented sensitivity, dynamic range, broadband spectral coverage with high spectral, temporal, and spatial resolution, high fidelity, and wideband imaging spectropolarimetry across decimetric–metric-centimetre  \citep{nindos2019}. With its capability for high-cadence, wideband imaging spectropolarimetry, SKA will enable systematic estimation of coronal magnetic fields, plasma turbulence, particle acceleration, and shock dynamics using radio bursts, with a level of precision previously unattainable. This information is highly important, not only for our better understanding of the physics of solar phenomena, but also for being employed as boundary conditions and validation means for different models addressing processes at both kinetic and magnetohydrodynamic (MHD) scales.

In this review, we introduce the general phenomenology of solar radio bursts, summarise the state of current radio observations from ground- and space-based instruments, and highlight the limitations imposed by sensitivity, dynamic range, and sparse sampling of the corona. We then discuss in detail how SKA, through its widespread baseline distribution, accurate polarization calibration, and extremely dense instantaneous u–v coverage, will overcome many of these challenges. Particular emphasis is placed on SKA’s potential to image radio bursts and their fine structures, quantify coronal magnetic topology via gyroresonance and gyrosynchrotron diagnostics and polarization observations of plasma emission, trace shock evolution and SEP source regions with unprecedented accuracy, and enable synergistic science with solar missions such as Solar Orbiter \citep[SolO;][]{muller2020}, Parker Solar Probe \citep[PSP;][]{psp2016}, Aditya-L1 \citep{Tripathi2023AdityaL1} and PROBA3 \citep{Zhukov2025ASPIICS}. Together, these capabilities position SKA as the key facility for the next generation of solar and heliospheric radio science. 

\section{Solar Radio Observational Techniques}

In observations of solar radio emission, we distinguish four different techniques, resulting in four distinct types of radio data: single-frequency measurements, dynamic spectra, radio imaging, and direction-finding observations.

\subsection{Single Frequency Observations}
\label{sec:section2.1}
\begin{figure}[!ht]
    \centering
	\includegraphics[width=\columnwidth]{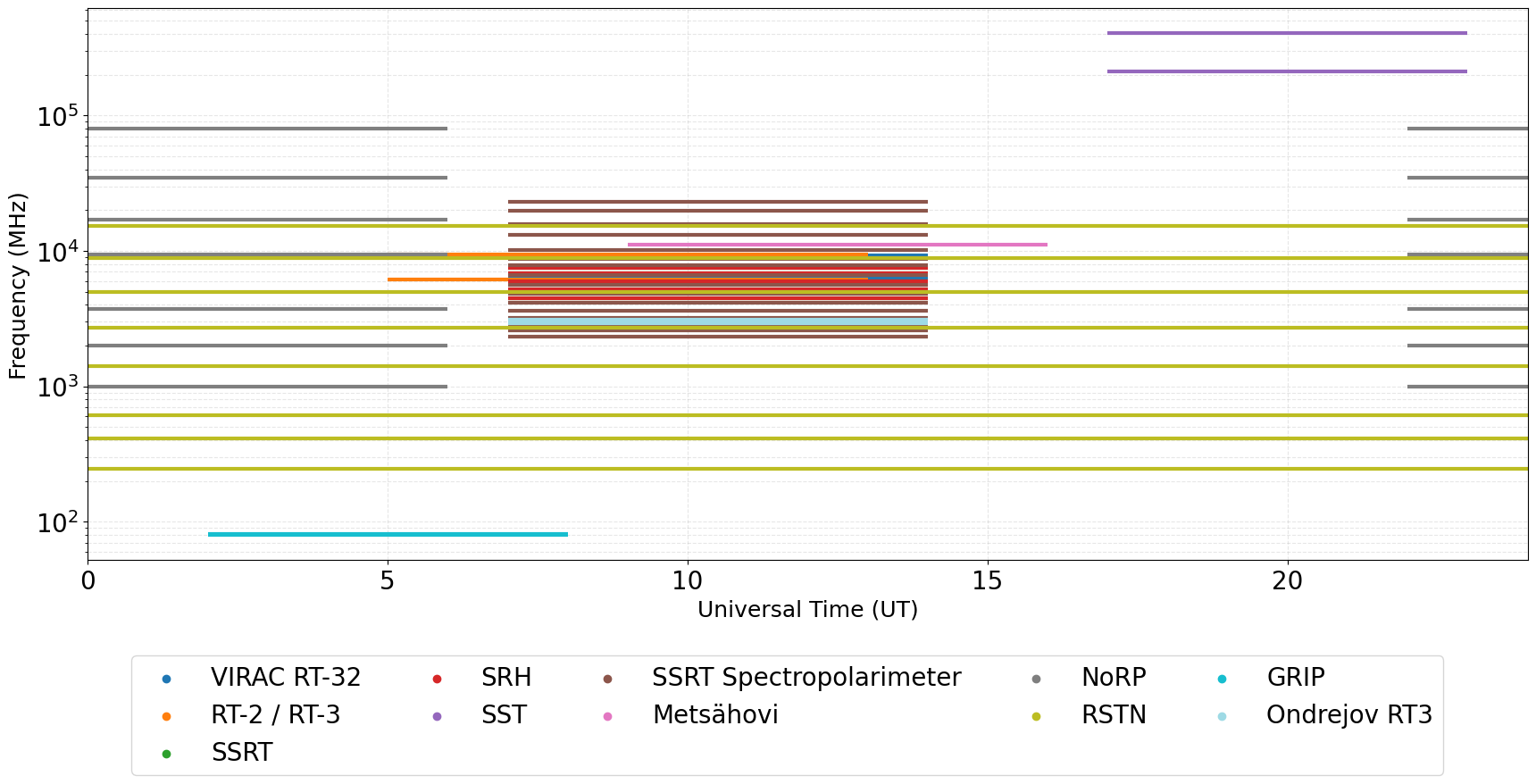}
    \caption{Daily observing windows and operational frequency ranges of non-imaging solar radio flux and light-curve instruments. Each horizontal bar represents the combined time–frequency coverage for a specific instrument, plotted in UT (horizontal axis) and frequency in MHz (vertical axis, logarithmic scale). These instruments measure integrated solar radio flux at one or multiple discrete frequencies, providing high-cadence time series used to study flare energetics, emission mechanisms, event timing, and long-term solar radio variability. Instruments with observing sessions crossing midnight (e.g., NoRP) are represented with two split rectangles. Together, these facilities offer continuous global monitoring across 1–400 GHz, which is crucial for characterizing impulsive and slowly varying solar radio emissions. }
     \label{fig:figure1}
\end{figure}

From the beginning of the radio observations until about 1950, all radio observations were based on single-frequency measurements that recorded both the continuum emission and radio bursts in the received radiation flux. Single-frequency data are radio polarimeter recordings with a relative frequency bandwidth typically ranging around df/f = 0.2 per cent. The currently operational single–frequency solar radio observatories, together with their frequency ranges and typical observing times, are summarised in Figure~\ref{fig:figure1}.

Currently, single-frequency radio observations are not very frequent, for example, the Gauribidanur Radio SpectroPolarimeter (GRIP) \citep{2013ApJ...762...89R} and the Radio Solar Telescope Network (RSTN) \citep{2022RaSc...5707456G}. The RSTN is operated by the United States Air Force and consists of four ground-based solar radio observatories—Sagamore Hill, Palehua, Learmonth, and San Vito—collectively known as the Radio Solar Telescope Network (RSTN). Each site continuously monitors solar radio emission at eight fixed frequencies (245, 410, 610, 1415, 2695, 4995, 8800, and 15400~MHz).
The Learmonth Solar Observatory \citep{Kennewell1983_Learmonth} is jointly operated by the Bureau of Meteorology (Australia) and the USAF.
The Ventspils International Radio Astronomy Center \citep[VIRAC;][]{bezrukovs2012receiving} RT-32 in Irbene, Latvia, performs single-dish test observations in the 6.3–9.3~GHz range, offering measurements in 16 frequency bands in both right circular polarization (RCP) and left circular polarization (LCP). At the Kislovodsk Mountain Astronomical Station of the Pulkovo Observatory (Russia), the RT-2 and RT-3 antennas operate at 6.15 and 9.35 GHz, with routine observations conducted between 05:00 and 13:00 UT. Real-time data have been available since April 2002, with archival observations dating back to 1957. The Siberian Solar Radio Telescope \citep[SSRT;][]{Smolkov1986SSRT} in Irkutsk provides correlation curves at 4.5, 5.2, 6.0, 6.8, and 7.5 GHz, while the Siberian Radioheliograph \citep[SRH;][]{Lesovoi2017SRH}, also located in Irkutsk, supplies correlation plots (Stokes I and V) at the same frequencies, with continuous observations commencing in July 2016. The Solar Submillimeter-wave Telescope \citep[SST;][]{Kaufmann2008SST} at El Leoncito, Argentina, measures full-Sun emission at 212 and 405~GHz using a cluster of six beams (four with HPBW $\sim$4$'$ at 212~GHz and two with HPBW $\sim$2$'$ at 405~GHz). The SSRT spectropolarimeter, operating between 2–24~GHz across 16 discrete channels (2.34–22.93~GHz), provides daily dual-polarization (RCP/LCP) data since March 2011. The Metsähovi Radio Observatory (MRO) in Finland observes the Sun at 11.2 GHz with a 1 kHz sampling rate, recording data continuously from sunrise to sunset since 2001. The Ond\v{r}ejov Solar Radio Observatory (Czech Republic) operates a single–frequency solar radio monitoring system at 3\,GHz \footnote{\url{https://space.asu.cas.cz/~radio/}}.
The Nobeyama Radio Polarimeters (NoRP) in Japan monitor full-disk solar emission at 1, 2, 3.75, 9.4, 17, 35, and 80 GHz with a time resolution of 0.1 s, providing one of the longest and most frequently used high-cadence datasets for solar activity studies \citep{Nakajima1985Nobeyama}.

\begin{figure}[!ht]
    \centering
	\includegraphics[width=0.8\columnwidth]{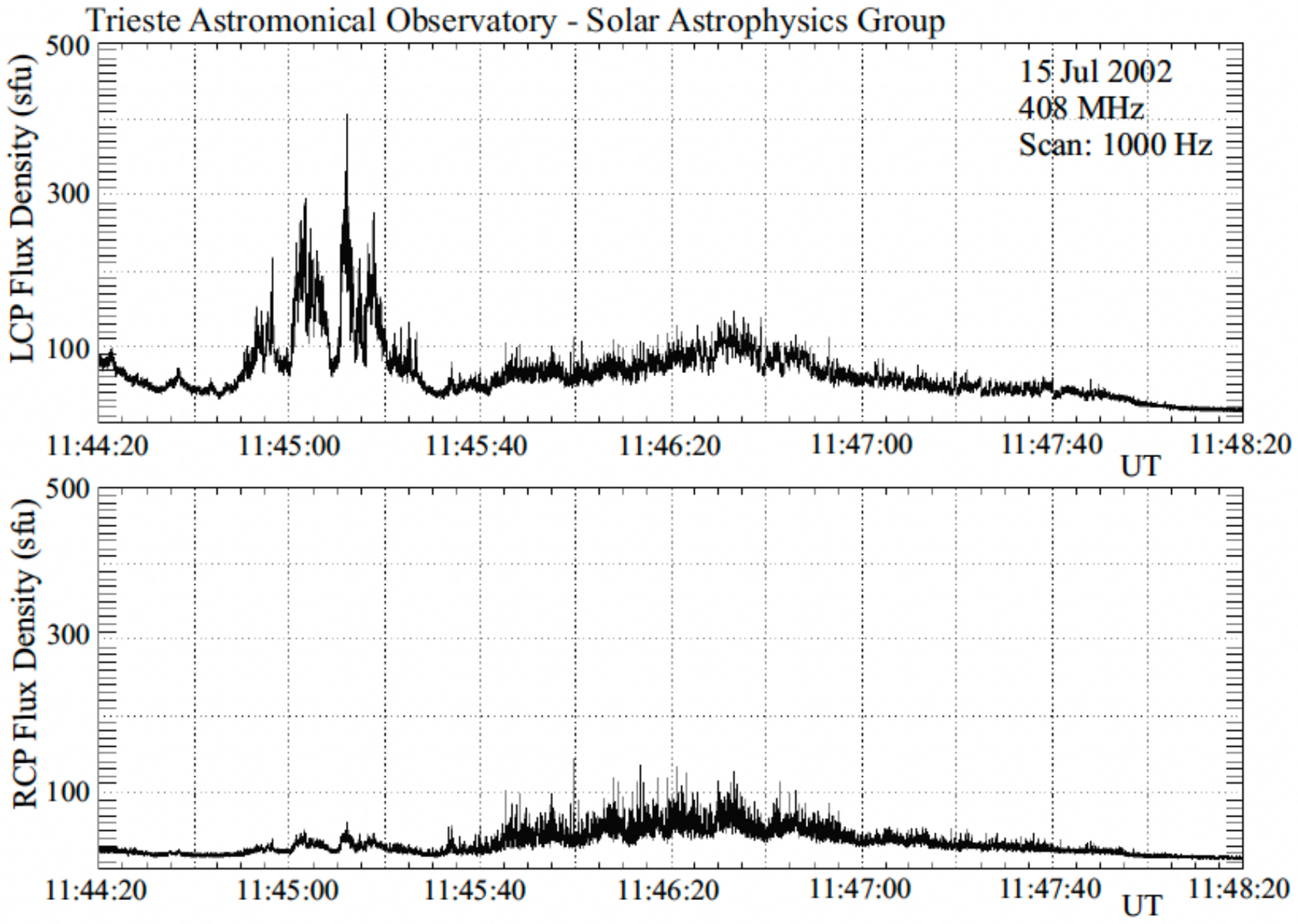}
    \caption{Single frequency data, i.e. time series of the flux density in sfu, recorded by the solar multichannel radiopolarimeter of the TSRS. The flux density, i.e., the L-hand and R-hand circular polarization measurements (top and bottom panel, respectively), were routinely taken in the metric range. In the first 80s of recordings, the structured continuum is strongly L-polarized (about 90~\%), and the continuum in the following 160s is weakly L-polarized. (Modified from \cite{Magdalenic2008Thesis}). }
     \label{fig:figure2}
\end{figure}

The example of fluctuations of the solar radio emission at a single frequency is shown in Figure~\ref{fig:figure2}. Presented observations were taken by the solar multichannel radio polarimeter of the Trieste Solar Radio System \citep[TSRS;][]{Messerotti2001} of the INAF-Trieste Astronomical Observatory. Observations of the integrated full disk emission were taken at a high temporal resolution, i.e., a sampling rate of 1000 Hz (1 ms). The TSRS instrument also provided very important additional information on the polarization of the recorded radiation, i.e., measurements of the flux density in LCP and RCP. The radio energy measured at the Earth, i.e. flux density, is in solar flux units sfu, where 1~sfu=$10^{−22}~W~m^{−2}~Hz^{−1}$. The TSRS instrument is unfortunately no longer operational. It should be noted that single-frequency observations are often strongly affected by the interference from local TV, radio emissions and other terrestrial radio sources. 

One of the most frequently used single frequencies, which is also considered a proxy for solar activity, is the so-called 10.7cm flux observations. Continuous monitoring of the solar radio flux at microwave frequencies (typically near 2.8~GHz) is one of the most reliable long-term indicators of solar magnetic activity. The 10.7 cm (2.8 GHz) solar flux index, commonly referred to as the F10.7 flux, has been routinely measured since 1947 at the Dominion Radio Astrophysical Observatory (DRAO) in Canada, providing an uninterrupted record spanning over seven decades. This index is highly correlated with the evolution of active regions, sunspot numbers, and extreme ultraviolet (EUV) irradiance, making it an essential proxy for tracking solar cycle variability. Complementary measurements at neighbouring frequencies, such as the 2.7 GHz observations from the Learmonth Solar Observatory in Australia, extend and support global monitoring efforts, ensuring continuity in the event of local interruptions or adverse weather conditions. Because the F10.7 flux responds sensitively to changes in the solar chromosphere and low corona, it is widely used in operational space-weather models, thermospheric–ionospheric density predictions, and satellite drag calculations. Despite its apparent simplicity as a single-frequency radio measurement, the F10.7 index remains one of the most robust and extensively utilised empirical parameters in heliophysics, linking solar activity to geospace response. With the continuously increasing number of radio interference sources, F10.7 flux observations, even in the 'protected frequency band', are being significantly endangered and are becoming less frequent.

\subsection{Dynamic Spectrometer and Spectropolarimeters}
\label{sec:section2.2}

\begin{figure}[!ht]
    \centering
	\includegraphics[width=\textwidth]{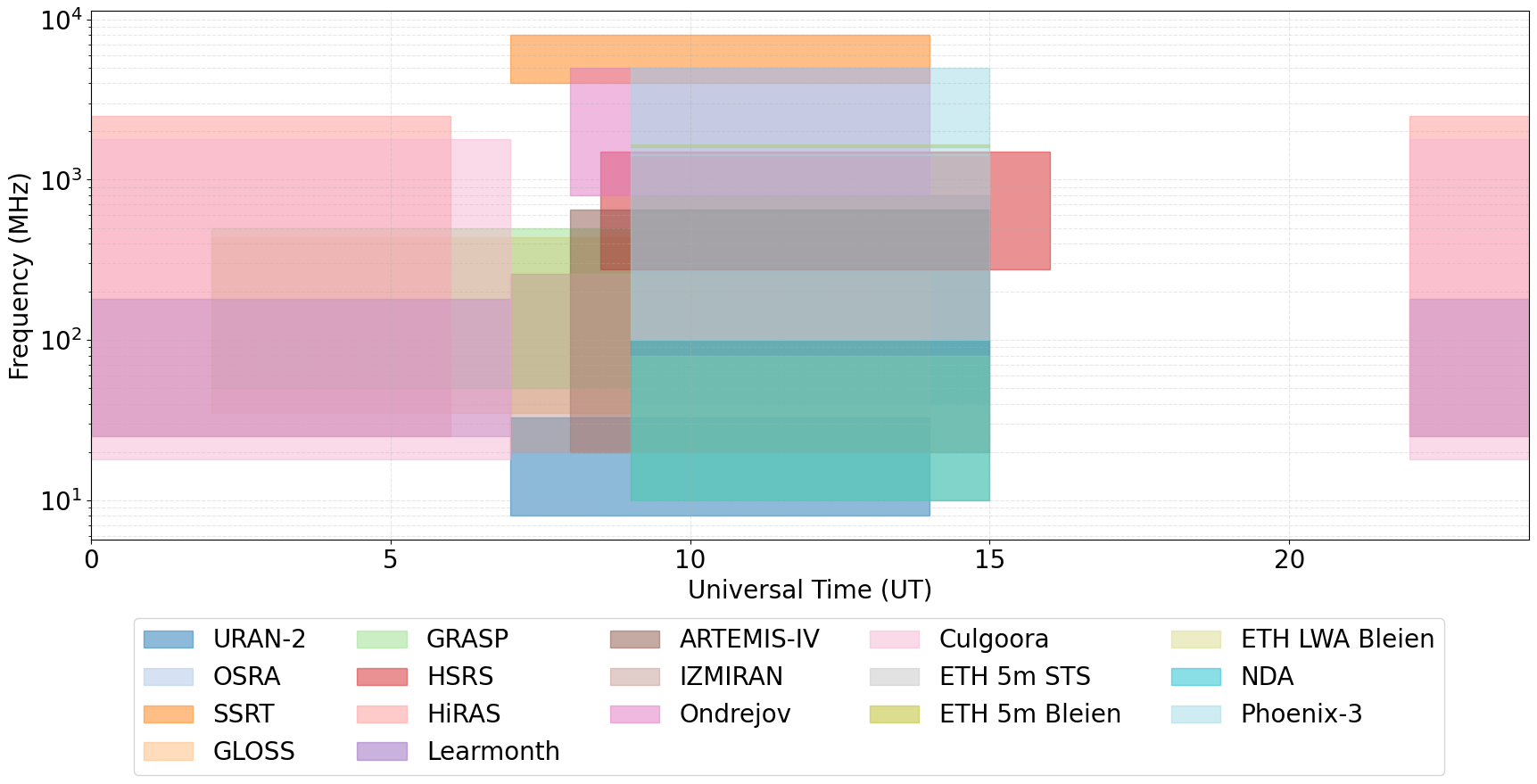}
    \caption{Operating time windows and frequency coverage for major solar dynamic spectrometers and spectropolarimeters that record full-Sun radio spectra with fine spectral and temporal resolution. These instruments produce dynamic spectra essential for identifying and classifying solar radio bursts (Types I–V), diagnosing electron acceleration, tracing the propagation of CMEs/shocks, and probing coronal/heliospheric plasma. Frequency ranges span from long-wavelength decametric systems (e.g., URAN-2 at 8–33 MHz) to microwave spectrometers (e.g., SSRT 4–8 GHz) and high-frequency broadband systems such as Phoenix-3 (100–5000 MHz). Many facilities operate only during local daytime (e.g., Humain, Trieste, Gauribidanur), while others maintain seasonal schedules (e.g., SSRT spectrograph). Instruments are shown excluding the distributed CALLISTO network, focusing instead on dedicated high-performance spectrometers with long-term archives. This global ensemble provides near-continuous spectral coverage from 10 MHz to several gigahertz, enabling comprehensive studies of solar radio bursts.}
     \label{fig:figure3}
\end{figure}

\begin{figure}[!ht]
    \centering
	\includegraphics[width=0.8\columnwidth]{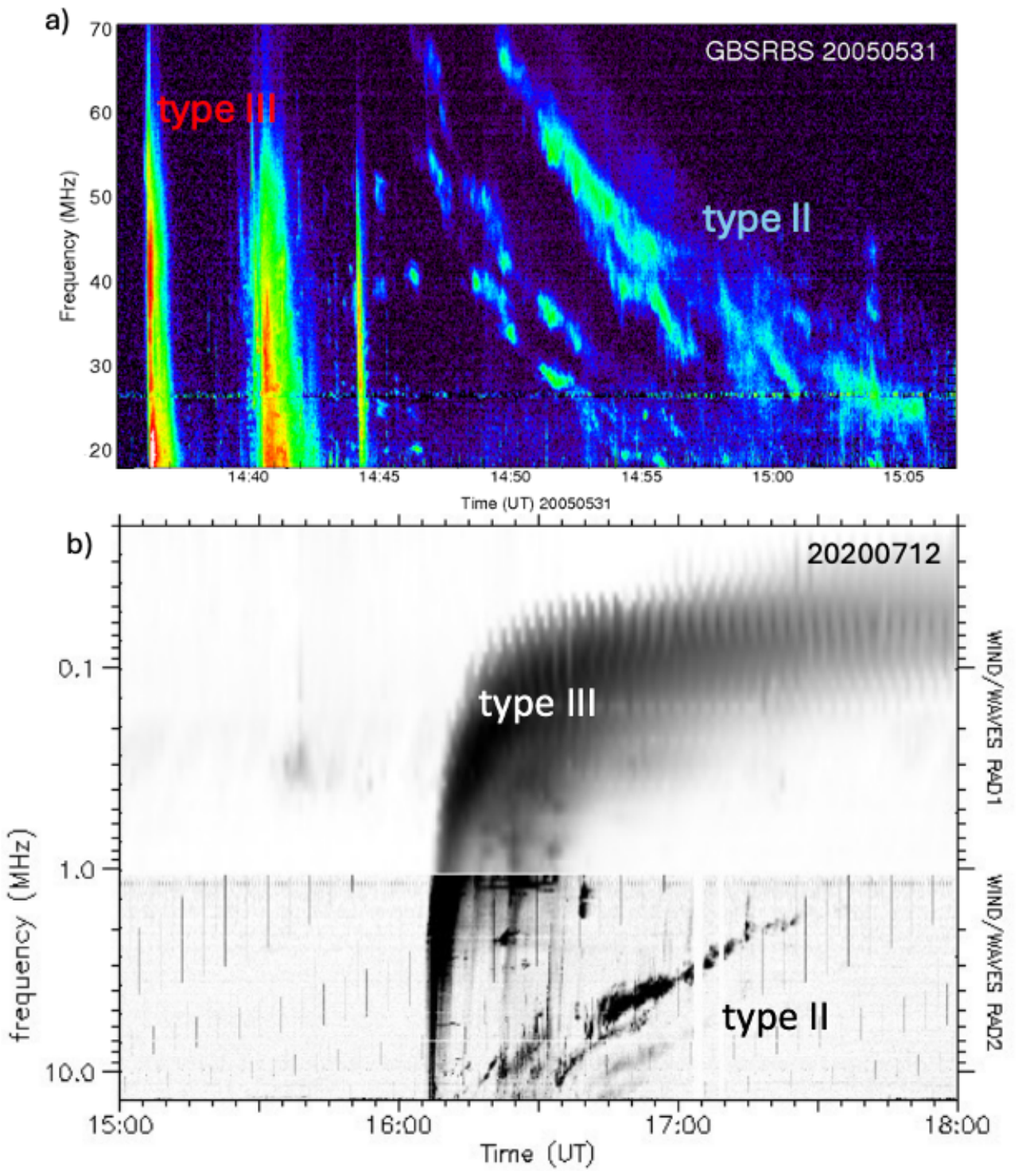}
    \caption{Dynamic spectra observed by ground-based and space-based instruments, panels a) and b) respectively. The slowly drifting type II radio bursts, i.e. signatures of shock waves and fast drifting type III radio bursts, i.e. signatures of fast electron beams, are marked in two spectra. a) Dynamic radio spectrum observed by Green Bank Solar Radio Burst Spectrometer in the frequency range 70–18 MHz, observed on May 31, 2005. 
b)  The radio emission in DH to km wavelength range observed by WIND/Waves instrument \citep{bougeret1995} on December 7, 2020.}
     \label{fig:figure4}
\end{figure}

Using only single-frequency measurements often makes it difficult to distinguish between different types of radio bursts, and additionally, the information about the frequency span of the bursts is lacking. \cite{Wild50a} introduced the technique of acquiring the dynamic spectra of radio bursts where the intensities are recorded continuously, as a function of frequency over a given frequency range. The radio spectrogram, also known as a dynamic radio spectrum, is a graphical representation of the radio emission intensity recorded at a number of closely spaced single frequencies as a function of time. The currently operational spectrometers and spectropolarimeters for solar radio observatories, together with their frequency ranges and typical observing times, are summarised in Figure~\ref{fig:figure3}. Dynamic spectra are often defined as frequency-time diagrams with colour-coded intensity of the radio emission.  Following the convention, the observing frequency in the metric wavelength range is presented, decreasing from top to bottom in Figure~\ref{fig:figure4}. The dynamic spectra recorded at longer wavelengths by the space-based radio instrument are generally presented with the frequency decreasing from the bottom to the top, opposite to the metric range convention (Figure ~\ref{fig:figure4}).
Dynamic spectra can be formed in several different ways. The previously often employed receivers, which “sweep” through the frequencies, measure radio emission at one frequency for a short time interval and then at the next instant at the following higher/lower frequency.
The metric range observations are often impacted by the radio frequency interferences (RFIs). Horizontal, uniformly coloured stripes (mostly white), often seen in dynamic spectra extending across the entire plot, are artificial emissions, i.e., interference from sources such as TV and radio stations. 

Whole-sun dynamic spectra are provided by several dedicated broadband radio spectrographs operating worldwide. 
The URAN--2 system in Poltava, Ukraine, observes in the 8--33 MHz range with dual polarization, a time resolution of 1--100 ms, and a frequency resolution of 4 kHz, offering a dynamic range of about 90 dB. 
The Observatory for Solar Radio Astronomy (OSRA) at Tremsdorf (Germany) operated between 40--800 MHz from 1954 to 2007, providing one of the longest-running spectral datasets. 
The Siberian Solar Radio Telescope (SSRT) in Irkutsk (Russia) supplies spectral observations in the 4--8 GHz range with 26 frequency channels and dual circular polarization; 1-s cadence data are publicly available, while 10-ms resolution data can be requested for detailed studies. 
The Gauribidanur Low-Frequency Solar Spectrograph (GLOSS) in India routinely monitors the 35--85 MHz range (recently upgraded to 35--435 MHz) with a cadence of 0.25 s, providing daily spectra and an extensive archive of type II bursts since 2009\footnote{\url{https://www.iiap.res.in/centers/gro/grids}}. The Gauribidanur RAdio Solar Spectropolarimeter (GRASP) in India routinely monitors the 50--500 MHz range with a cadence of 0.25 s, providing daily spectra in Stokes-I and Stokes-V \citep{Kumari2015}.
High-resolution observations at decimetric and metric wavelengths are provided by the Humain Solar Radio Spectrograph (HSRS) in Belgium, operating between 275--1495 MHz with roughly 12{,}500 channels and a cadence of 0.25 s \footnote{\url{https://www.sidc.be/humain/home}}. 
The Hiraiso Radio Spectrograph (HiRAS) in Japan spans an exceptionally wide band of 25--2500 MHz and has been delivering continuous solar dynamic spectra since 1996 \footnote{\url{https://solarobs.nict.go.jp/radio.html}}. 
The Learmonth Solar Radio Spectrograph (Australia) sweeps through 25--180 MHz every three seconds and remains an important component of long-term solar monitoring. Ond\v{r}ejov Solar Radio Telescopes (Czech Republic) provide detailed dynamic spectra in the 2--4.5 GHz range, which are widely used for studying coherent microwave bursts. 
Additional low-frequency dynamic spectra have historically been supplied by the Green Bank Solar Radio Burst Spectrometer \citep[GBSRBS;][]{White2005} in the USA  (see Figure ~\ref{fig:figure4}), which operated from 18 to 70 MHz until 2012 \citep[GBSRBS;][]{White2005}, and by the Culgoora Solar Radio Spectrograph (Australia), which surveyed 18 to 1800 MHz with a three-second sweep cadence for several decades. 
Complementary measurements at centimetric wavelengths are obtained using the ETH Zurich 5-m dish in Bleien, Switzerland, which observes between 1590 and 1650 MHz with a temporal resolution of 0.25 s. 
At lower frequencies, the ETH Zurich LWA system at Bleien covers 10--80 MHz with dual circular polarization, 0.25-s cadence, and a 300-kHz bandwidth. 
Finally, the Phoenix-3 spectrograph (Bleien, Switzerland) provides high-quality dynamic spectra from 100 to 5000 MHz at a sub-second cadence ($\approx$0.1 s) with 5000 spectral channels, offering one of the most detailed broadband datasets currently available. 
The e-CALLISTO network is a globally distributed set of low-cost, programmable spectrometers designed for continuous monitoring of solar radio emission in the 45–870~MHz range. Built around the CALLISTO receiver concept developed at ETH Zurich \citep{be09b}, the network provides near-24-hour coverage of solar radio bursts and RFI environments, with up to 400 channels per sweep and a typical time resolution of 0.25 seconds, making it a crucial resource for heliophysics research, education, and space weather monitoring \footnote{\url{https://www.e-callisto.org/Data/data.html}}.

The dynamic radio spectra are also affected by the radio emissions of terrestrial origin. However, due to the continuous and broadband mapping of radio emission along the range of frequencies, the effect is somewhat smaller than in the case of single-frequency observations.

\subsection{Solar Radio Imaging}
\label{sec:section2.3}

Imaging observations of the solar corona at radio wavelengths are essential for locating emission sources, tracking coronal dynamics, and linking radio bursts to underlying plasma processes. 
One of the most important radio instruments, which has provided solar radio images for several decades and whose observations have resulted in numerous publications and several breakthroughs in solar radio physics, is the Nançay Radioheliograph \citep[NRH;] []{ke97}. The NRH instrument is located in France and consists of two linear, cross-shaped antenna arrays: an East-West arm with 19 antennas distributed along a 3,200 m baseline, and a North-South arm with 24 antennas spanning 1,250 m. Together, these provide 576 interferometric visibilities. The instrument observes presently at ten fixed frequencies between 150 and 450 MHz (150.9, 228.0, 298.7, 382.2, 432.0, 173.2, 270.6, 327.0, 408.0, and 444.0 MHz), enabling true two-dimensional snapshot imaging of the solar corona at a cadence of up to 8 images per second. With its digital correlator and dense instantaneous \emph{uv}-coverage, the NRH routinely delivers high-quality images of metric-wavelength emission, including all types of radio bursts and their substructures. 
NRH data are publicly available\footnote{\url{http://bass2000.obspm.fr/home.php}}. 

\begin{figure}[!ht]
    \centering
	\includegraphics[width=\textwidth]{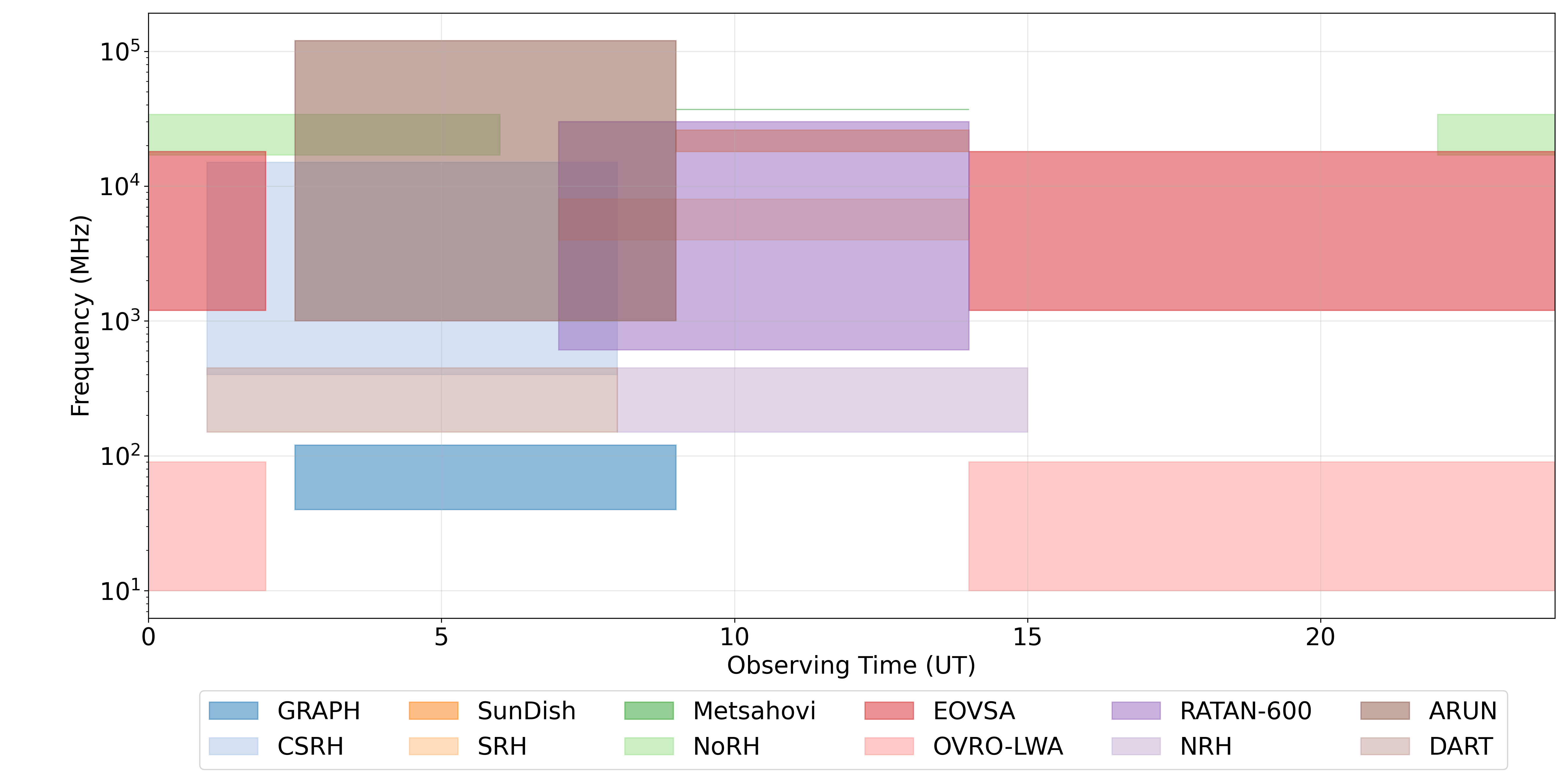}
    \caption{Time–frequency coverage for solar-dedicated radio interferometers and imaging arrays, plotted across UT and operating frequencies. These instruments are designed specifically for high-resolution imaging of the solar corona, mapping active regions, flares, CMEs, and coronal magnetic field topology. Shown are facilities such as GRAPH (40–120 MHz), CSRH / MUSER (0.4–15 GHz), SunDish (18–26 GHz), Siberian Radioheliograph (4–8 GHz), Metsähovi 37 GHz solar system, NoRH (17 \& 34 GHz), EOVSA (1–18 GHz), OVRO-LWA, NRH (150–450 MHz), RATAN-600, ARUN (1–12 GHz solar imager) and others. Instruments crossing midnight (e.g., NoRH, EOVSA) are represented with wrap-around blocks. Collectively, these systems provide imaging across 0.04–37 GHz, offering unprecedented diagnostics of coronal magnetic fields, source heights, electron populations, and plasma dynamics from metric to millimeter wavelengths.}
     \label{fig:figure5}
\end{figure}

Together with NRH, several other instruments situated worldwide provide complementary solar radio imaging at microwave, decimetric, and metric wavelengths. Herein, we mention some of them. The SunDish project in Italy (Cagliari and Bologna) performs 2D spectropolarimetric imaging between 18 and 26 GHz with an angular resolution of approximately 1 arcmin, with weekly observations since 2018 \citep{SunDishProject}. The Siberian Solar Radio Telescope \citep[SSRT; see e.g.][]{Grechnev2003} in Irkutsk, Russia, operates at 5.7~GHz and has been producing two-dimensional solar maps since 1997, making it one of the longest-running high-frequency imaging facilities. At metric wavelengths, the NRH (described above) provides dense snapshot imaging between 150 and 450 MHz using a 47-antenna interferometric array operating at up to ten frequencies. In India, the Gauribidanur Radioheliograph \citep[GRAPH;][]{1998SoPh..181..439R} observes at $40-120$ MHz with a cadence of 0.25~s, generating daily solar images during local meridian transit (04:00--09:00~UT). Continuous observations have been available online since 2014. GRAPH is particularly valuable for tracking low-coronal Type~III bursts, noise storms, and CME-associated sources at long wavelengths. The Daocheng Radio Telescope (DART) is a 1-km diameter, 313-element circular interferometric array operating at 150–450 MHz, designed for high-fidelity solar coronal imaging and complementary low-frequency radio astronomy observations in China. The Chinese Spectral Radioheliograph (CSRH)/ Mingantu Spectral Radio Heliograph \citep[MUSER;][]{Yan2021}(MUSER; Yan et al., 2021) operates from 0.4-15 GHz.
The currently operational solar radio imaging instruments, together with their frequency ranges and typical observing times, are summarised in Figure~\ref{fig:figure5}.

At millimetric wavelengths, the Metsähovi Radio Observatory \citep[MRO;][]{MRO_solar_database} in Finland observes at 37 GHz, providing daily full-sun maps since 1978. During the summer months, extended observing sessions of up to 14 hours per day are conducted, with each map requiring approximately 150 seconds of integration time. Microwave imaging at 17 and 34~GHz is performed by Nobeyama Radioheliograph (NoRH). NoRH has been decommissioned since 2020. An upcoming solar radio instrument in India, the Advanced Radio Telescope Udaipur Network for Solar and Space Weather Research (ARUN--SSW), operating in the 1--12 GHz range, is planned to be commissioned in the coming years \footnote{\url{https://agu.confex.com/agu/agu25/meetingapp.cgi/Paper/1905282}}.
NoRH was delivering high-cadence (0.1--1\,s) imaging with spatial resolutions of 
$\sim 10''$ at 17\,GHz and $\sim 5''$ at 34\,GHz, 
including full Stokes~I and~V at 17\,GHz and Stokes~I at 34\,GHz. Its continuous daily operation since 1992 has produced an unparalleled archive of flare and active-region microwave images.

\begin{figure}[!ht]
    \centering
	\includegraphics[width=\textwidth]{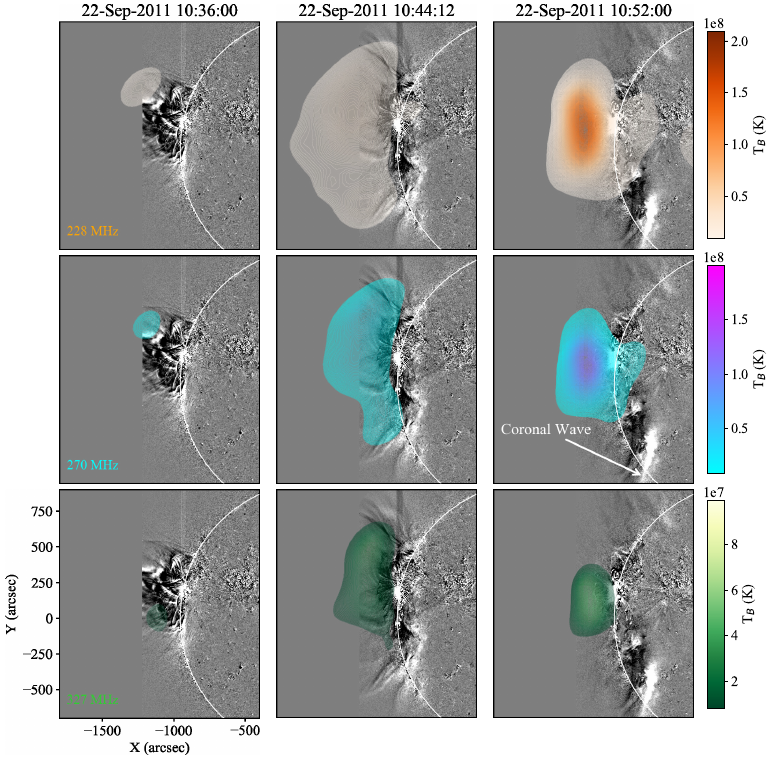}
    \caption{The radio sources from NRH (filled contours), during a type IV burst, overlaid over AIA 211 Å running-difference images. This figure is adapted from \citep{morosan2019variable}.}
     \label{fig:figure6}
\end{figure}

In the United States, the Owens Valley Solar Array \citep[EOVSA][]{Gary2018_EOVSA} offers imaging capability across 39 frequencies between 1.2 and 18 GHz, enabling broadband mapping of gyrosynchrotron emission from flaring regions. EOVSA offers a raw frequency resolution of 122 kHz, corresponding to 4096 spectral channels, while the effective science-grade resolution is approximately 50~MHz \footnote{For EOVSA and OVRO-LWA, the effective observing windows are approximately 14:00–02:00 UT during summer and 16:00–23:30 UT during winter.}.
The temporal sampling of the instrument is 20~ms, with a full spectral sweep completed every 1~s.  Similarly, the RATAN-600 telescope in Russia \citep{Korolkov1979} provides one-dimensional scans at 112 frequencies spanning 0.7--18.2~GHz. Observations are typically carried out between 07:00 and 11:00~UT, producing 5--60 scans per day, with full Stokes I and V recorded at all channels. Together, these imaging facilities provide multi-frequency, multi-resolution views of the solar corona, enabling the detailed study of the flare energy release process and electron acceleration in the dense coronal plasma. Figure~\ref {fig:figure6} shows the solar radio images in the metric range made with NRH.

Several powerful radio interferometers, although not dedicated solar instruments, regularly contribute to solar imaging through targeted campaigns or the dynamic allocation of observing time. The upgraded Giant Metrewave Radio Telescope \citep[uGMRT;][]{uGMRT2017} in India, operating in the range of 150-1450 MHz, has been extensively used for high–fidelity imaging of solar transients due to its large collecting area and flexible correlator modes. The LOw Frequency ARray \citep[LOFAR;][]{lofar2013} is another low frequency radio instrument. The LOFAR stations are spread across Europe, observing in the range 10--240 MHz, providing exceptional low–frequency coverage with dense core baselines suitable for coronal imaging. The Murchison Widefield Array \citep[MWA;][]{Tingay2013} in Australia, covering 80--300 MHz, routinely supports solar observations with snapshot uv–coverage that enables high–cadence imaging of rapidly evolving burst emission. At microwave frequencies, the Karl G. Jansky Very Large Array \citep[VLA;][]{Perley2011_vla} in the United States offers imaging from 1 to 50 GHz with full polarimetry, making it uniquely capable of probing gyrosynchrotron and gyroresonance sources in the low corona. MeerKAT \citep{jonas2016meerkat} in South Africa, operating between 0.6–14 GHz, provides excellent sensitivity and dynamic range that have recently enabled new solar imaging experiments.

\begin{figure}[!ht]
    \centering
	\includegraphics[width=0.8\columnwidth]{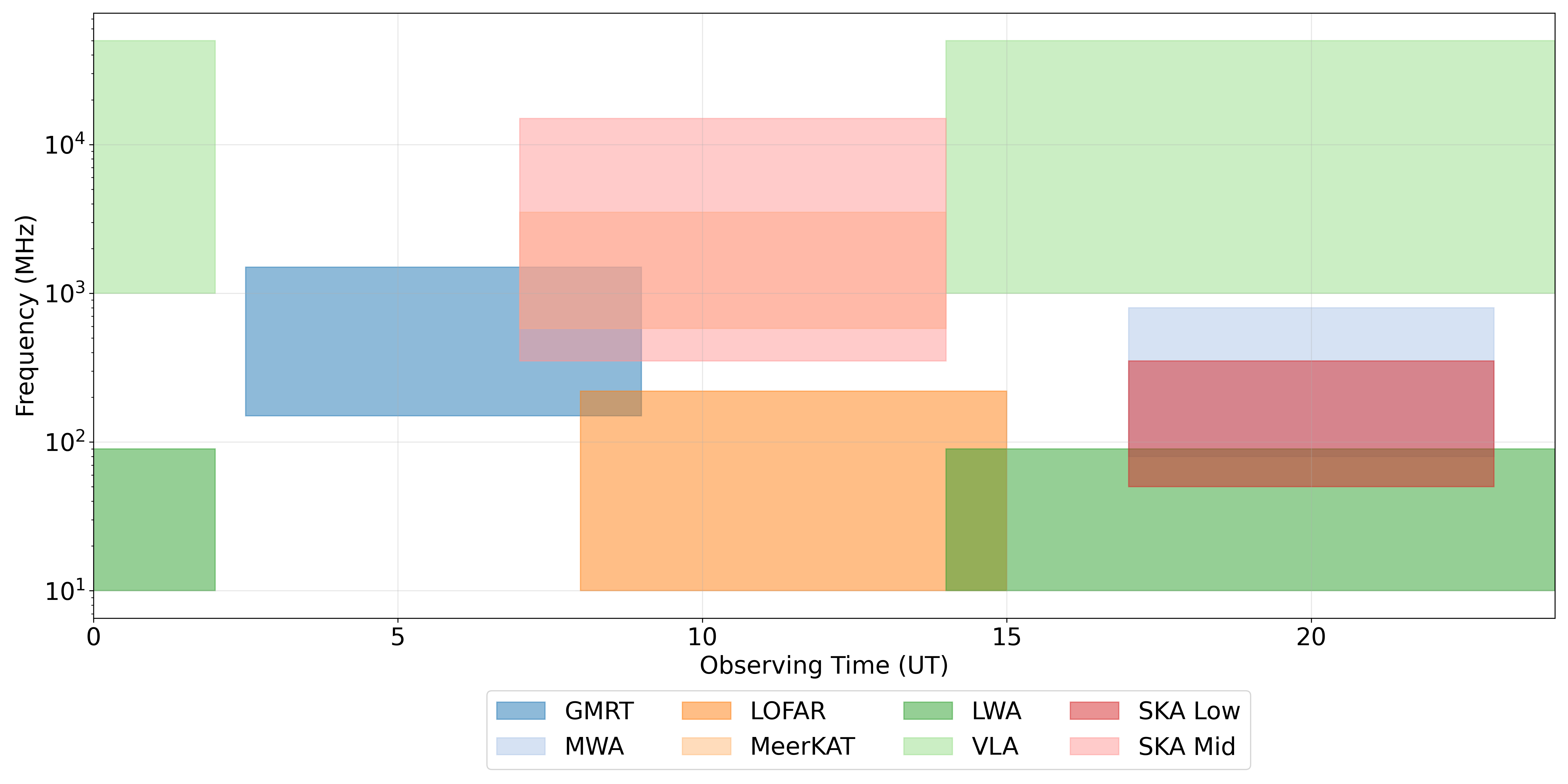}
    \caption{Observing windows and operational frequency bands for general-purpose radio interferometers that, while not solar-dedicated, routinely conduct solar imaging campaigns or support targeted observing modes. These include major low-frequency and microwave facilities such as GMRT, MWA, LOFAR, MeerKAT, LWA, VLA, SKA-Low, and SKA-Mid. Their wide instantaneous bandwidths and flexible correlator architectures enable high-dynamic-range imaging of the Sun on demand, though scheduling is constrained by competition with non-solar programs. Frequency ranges span 10 MHz to 50 GHz, allowing these arrays to capture a wide variety of solar radio phenomena from low-coronal plasma emission to gyrosynchrotron and gyroresonance processes. These instruments complement solar-dedicated arrays by providing superior sensitivity, large collecting areas, and advanced imaging fidelity, thereby extending the global solar radio observing capability into professional radio astronomy infrastructures.}
     \label{fig:figure8}
\end{figure}

Next–generation facilities will further expand these capabilities. The Long Wavelength Array (LWA; 10–88~MHz) in the United States provides ultra–low frequency imaging suited for coronal plasma processes \citep{2010iska.meetE..24H}, while the Square Kilometre Array (SKA)---comprising SKA-Low (50–350~MHz) in Australia and SKA-Mid (0.35–15.3~GHz) in South Africa \citep[see e.g.,][]{Henning10,nakariakov2015solar,dewdney2017mid} will deliver wideband, high–dynamic–range imaging spectropolarimetry across nearly two decades of frequency. Although these observatories prioritise non-solar astrophysical science, their large collecting areas, sophisticated correlators, and flexible scheduling make them invaluable contributors to solar radio research, complementing dedicated solar arrays and significantly extending the global imaging capability of the heliophysics community. The currently operational non-solar radio imaging instruments, occasionally used for solar observations, together with their frequency ranges and typical observing times, are summarised in Figure~\ref{fig:figure8}.

\subsection{Direction finding observations}
\label{sec:sec2.4}
\begin{figure}[ht]
    \centering
    \includegraphics[width=0.9\columnwidth]{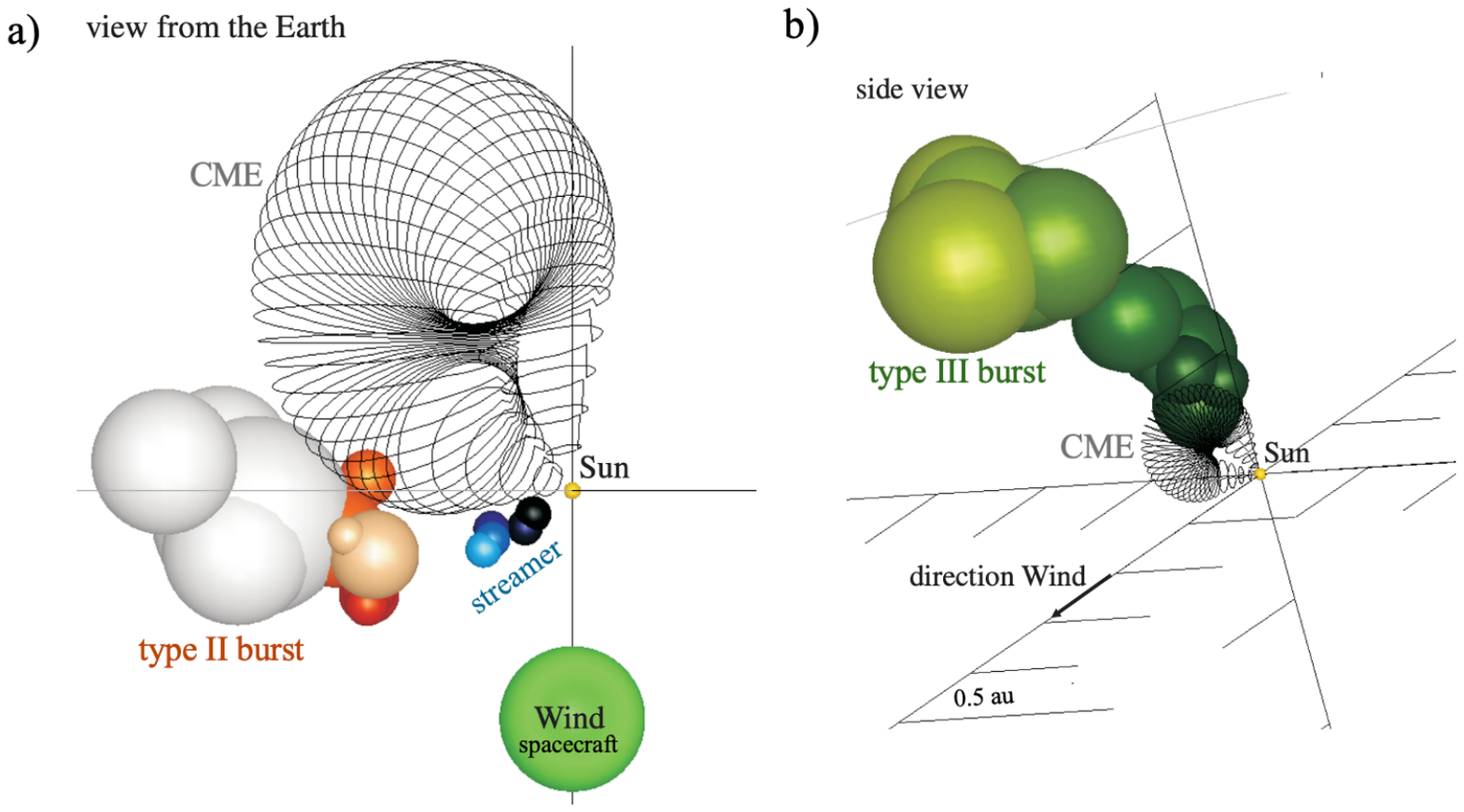}
    \caption{The propagation path of the type II and type III radio bursts obtained using radio triangulation (panels a and b), respectively. The studied radio event was associated with the CME/flare event observed on March 05, 2012. The size of the colourful spheres is half the distance between the two wave vectors obtained using the direction finding data. The sources are colour-coded, with the lighter colour corresponding to the lower observing frequency pairs. The black grid croissant represents the 3D reconstruction of the CME flux rope in the studied event, while blue spheres represent the 3D reconstruction of the nearby streamer. \citep[Figure is adapted from][]{magdalenic14}.}
    \label{fig:figure9}
\end{figure}

Mapping the positions of radio sources at frequencies below approximately 10 MHz is not possible with interferometric arrays, i.e., with ground-based instruments, due to the ionospheric cutoff. To estimate the locations of radio sources at large wavelengths and far from the Sun, one-dimensional (1D) electron density models \citep[e.g.,][]{new61, saito77, Leblanc95} have been employed for decades when no other observations or methods were available. The 1D density models assume radial symmetry, and without the spatial information, they can provide only radial distance of the radio source positions \citep[see e.g.][]{Knock05, Magdalenic2008, Costas20}. On the other hand, the direction-finding (DF) observations from space-based instruments available over the last few decades provide significantly more information on the source locations than only the radial distance of the source. The direction finding observations are the only way to obtain the position of the radio sources in the 3D space, without using additional assumptions or simplifications \citealp[see e.g.][]{Krupar2012, magdalenic14, jebaraj2020}. The propagation path of solar radio bursts obtained using radio triangulation is shown in Figure \ref{fig:figure9}.

Direction-finding observations are single-spacecraft measurements allowing us to determine the direction of arrival of an incident radio wave \citep{Cecconi2005}. Spacecraft equipped with radio antennas measure properties of incoming radio waves, such as intensity, polarization, and the direction of arrival. The direction is typically expressed in terms of colatitude and azimuth relative to the spacecraft frame. This allows us to estimate the wave's direction in space, i.e. a line of sight to the radio source. Depending on the type of spacecraft, i.e., spinning (Wind) or three-axis stabilized (STEREO, SolO, PSP), different techniques are used to estimate the direction finding parameters \citep{Fainberg1974, Cecconi2005, Krupar2012}. A single spacecraft can only provide the direction of the wave vector. In order to obtain accurate three-dimensional (3D) radio source positions, the triangulation method using DF measurements from two or more widely separated spacecraft needs to be employed. The intersection region of the two wave vectors determines the source location of the radio emission \citep{Reiner98, magdalenic14, Krupar2016, jebaraj2020, deshpande25}.  

\subsection{Space-based solar radio observatories}

As introduced in the previous section, in order to overcome the ionospheric cutoff, the frequencies below $\sim$10~MHz need to be observed by the space-based instruments. All types of solar radio bursts, except type I bursts, are observed at these low frequencies. They often, but not necessarily, start in the metric range and continue to the kilometric wavelengths. The radio bursts observed by space-based instruments provide important information on the characteristics of the interplanetary medium, which is essential in the validation of the space weather models and in the operational space weather forecasting \citep[see e.g.][and references therein]{magdalenic14, Valentino24, deshpande25}. 

The first ever solar radio observations by space-based instrument was performed by RAE-1 \citep[Radio Astronomy Explorer;][]{alexander1970rae1}. Dynamic spectra obtained by the radiometer spanned a frequency range of approximately 0.2 to 10 MHz. Since then the missions such RAE-2 \citep{alexander1975rae2}, Helios~1 and 2 \citep{porsche1977helios1, porsche1979helios2}, Voyager~1 and 2 \citep{stone1977voyager1, stone1979voyager2}, ISEE~1, 2 and 3 \citep{riedler1977isee12, ogilvie1978isee3}, and Ulysses \citep{meeks1990ulysses} largely contributed to the spectroscopic observations. 

\begin{table}[ht]
\centering
\footnotesize
\setlength{\tabcolsep}{4pt}
\renewcommand{\arraystretch}{1.2}
\caption{Summary of major space-based solar radio instruments from 1970 to present.}
\begin{tabular}{p{3cm} p{2cm} p{2cm} p{1.5cm} p{3cm}}
\hline
\textbf{Mission / Instrument} & 
\textbf{Freq. Range (MHz)} & 
\textbf{Measurement} & 
\textbf{Years} & 
\textbf{Notes} \\
\hline

\multicolumn{5}{l}{\textbf{Currently Operating}} \\
\hline
Wind/WAVES & 0.02--13.825 & Spectrometer + DF & 1994-- & Reference for IP bursts \\
STEREO-A/SWAVES & 0.0025--16 & Spectrometer + DF & 2006-- & Full GP capability \\
Parker Solar Probe/RFS & 0.01--19.2 & Radio + E-field + DF & 2018-- & Closest to Sun \\
Solar Orbiter/RPW & 0.0005--16 & Radio + DF & 2020-- & Tri-axial antennas \\
BepiColombo Mio/PWI & <0.01 & Plasma waves & 2018-- & LF electric field only \\

\hline
\multicolumn{5}{l}{\textbf{Historical Missions}} \\
\hline
STEREO-B/SWAVES & 0.0025--16 & Spectrometer + DF & 2006--14/16 & Limited post-recovery \\
Ulysses/URAP & 0.001--1 & Radio + plasma waves & 1990--2009 & High-latitude mission \\
ISEE-3 (ICE)/RAE & 0.03--2 & Solar radio & 1978--1997 & Early DF \\
Helios A/B & up to ~1 & Radio + plasma waves & 1974--1986 & Inner-heliosphere data \\
IMP-6/IMP-8 & 0.03--1.2 & Solar radio & 1970--2001 & Early space dynamic spectra \\
Voyager 1/2 PWS & $10^{-5}$--0.056 & Plasma waves & 1977-- & IP shocks, not bursts \\
\hline

\end{tabular}
\label{tab:table2}
\end{table}

Modern space-based observations have undergone significant improvements since these early missions. After its launch in 1994, Wind/Waves \citep{bougeret1995} provides continuous observations of the dynamic spectra from the L1 Lagrange point. It also offers thermal noise observations and DF measurements. STEREO/SWaves \citep{Kaiser2008, bougeret200B} consists of two identical spacecraft, STEREO-A and STEREO-B, offering multi-point observations around the Sun at approximately 1~au, with STEREO-B observations available until October 2014. In addition to its broadband dynamic spectra, STEREO/SWAVES provides DF measurements. The novel spectroscopic observations by Parker Solar Probe \citep[PSP;][]{psp2016}, Solar Orbiter \citep[SolO;][]{muller2020} and Bepi Colombo provide high frequency and time resolution observations. These novel missions explore the inner heliosphere from various vantage points near the Sun. They observe the Sun from close distances, providing an opportunity to study solar radio emissions near their source regions. BepiColombo, PSP, and Solar Orbiter also provide thermal noise measurements. Additionally, PSP and Solar Orbiter are equipped with DF capabilities. Table~\ref{tab:table2} presents a summary of the major space-based solar radio instruments operating from 1970 to the present, highlighting their frequency coverage, measurement capabilities, and mission context. While the DF measurements can be used to obtain 3D radio source positions \citep{Krupar14b, Krupar14a, magdalenic14, jebaraj2020, deshpande25} as mentioned in Section \ref{sec:sec2.4},  multi-spacecraft spectroscopic observations can provide 2D information, i.e., radio source trajectory \citep[see e.g.][]{Musset21, Badman22}.

\section{Solar Radio Observations}
         
\subsection{Active Sun observations}\label{active_sun_obs}
\subsubsection{Types of Solar Radio Bursts}

\begin{figure}[!ht]
    \centering
	\includegraphics[width=0.8\columnwidth]{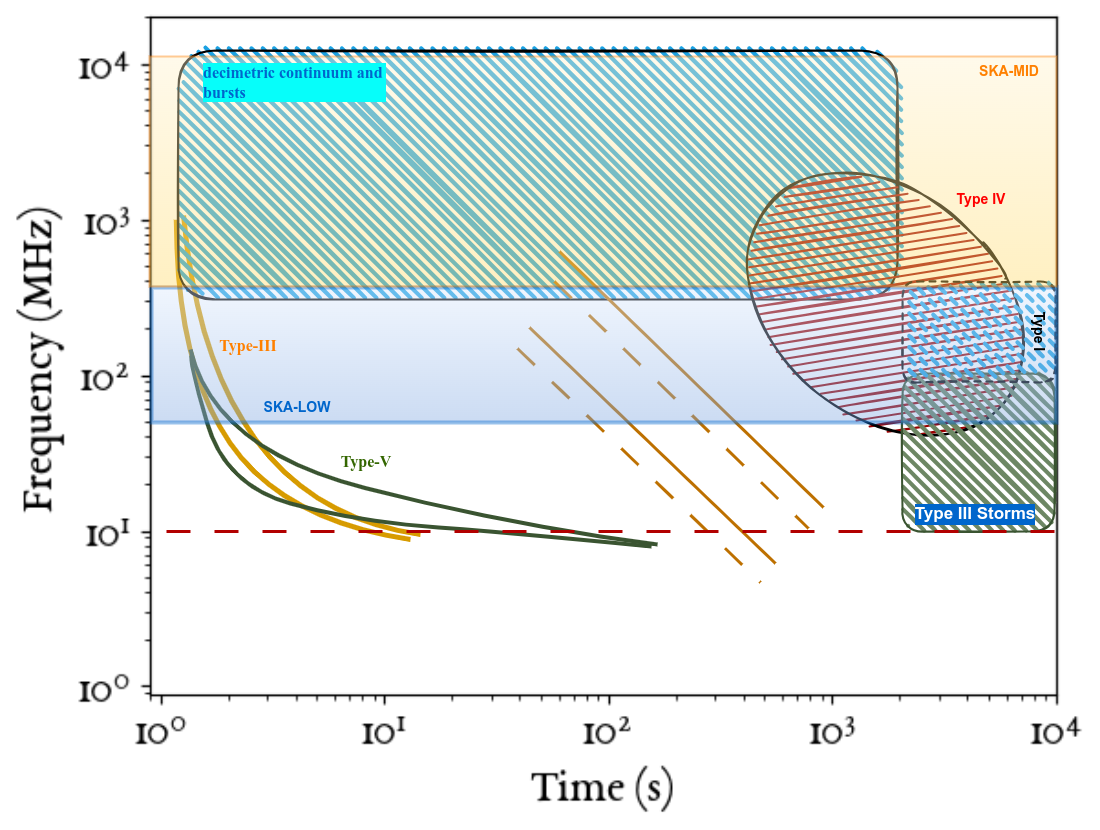}
        \caption{Different types of solar radio bursts and their characteristic signatures in the dynamic spectrum. Type~II bursts are slowly drifting emission lanes known as shock wave signatures. The rapidly drifting type III bursts are produced by very fast electron beams. Type~IV bursts present broadband, long-duration continuum often associated with the lift-up of the CME. Type~I noise storms appear as narrowband, short-lived bursts sometimes superimposed on a weak continuum background.}
    \label{fig:figure10}
\end{figure}

Solar radio bursts can be produced by energetic processes in the solar corona, such as solar flares and coronal mass ejections (CMEs), but they are also frequently observed in the absence of large eruptive events. These bursts occur following the acceleration of electron beams that can, in turn, generate emission at radio wavelengths most frequently through the plasma emission mechanism \cite[see e.g.][and the references therein]{kl02}. Solar radio bursts are classified based on their shape and characteristics in radio dynamic spectra as type I, II, III, IV and V \citep[][]{wild1950, Wild63, SuzukiDulk85}. Different types of solar radio bursts and their characteristic signatures in the dynamic spectrum are shown in Figure \ref{fig:figure10}.

\paragraph{Type I Bursts:}
Metric type I solar radio bursts are observed in dynamic spectra as narrowband (a few MHz), generally non-drifting bursts of short duration, typically about 0.1 s. When observed in single-frequency records, they appear as numerous “spiky” bursts of much higher intensity (several tens of sfu) than the intensity of the underlying background continuum. They usually occur in chains or small groups of bursts. Type I bursts are embedded in a weak, slowly varying continuum of emission, and the burst complex is referred to as a noise storm. In the metric-decametric wavelength range, type I bursts transition into type III bursts\citep{mugundhan2018spectropolarimetric}. The characteristics of individual bursts in the storm are generally similar to those found in chains or groups. Noise storms can last from a few tens of minutes to several days. Long duration is one of the main characteristics that distinguishes storms from other types of solar radio emissions (type II bursts or isolated type III bursts). Type I storms are generally associated with continuous reconnection above complex active regions \citep{Gary2004}, rather than directly with flaring activity. One possible mechanism for their generation is the avalanche model \citep{Mercier1997}. The frequency range in which they appear indicates that the emission mechanism is probably fundamental-frequency plasma emission. Some of the very important characteristics of type I storms, e.g., high brightness temperature and strong o-mode polarization (when the source is on the disk), can be explained by this hypothesis \citep[for review see e.g.][]{McLean1985}. 
It should be stressed that type I bursts are the most common type of solar radio bursts in metric wavelengths, but they are not very often studied.

\begin{figure}[!ht]
    \centering
	\includegraphics[width=0.8\columnwidth]{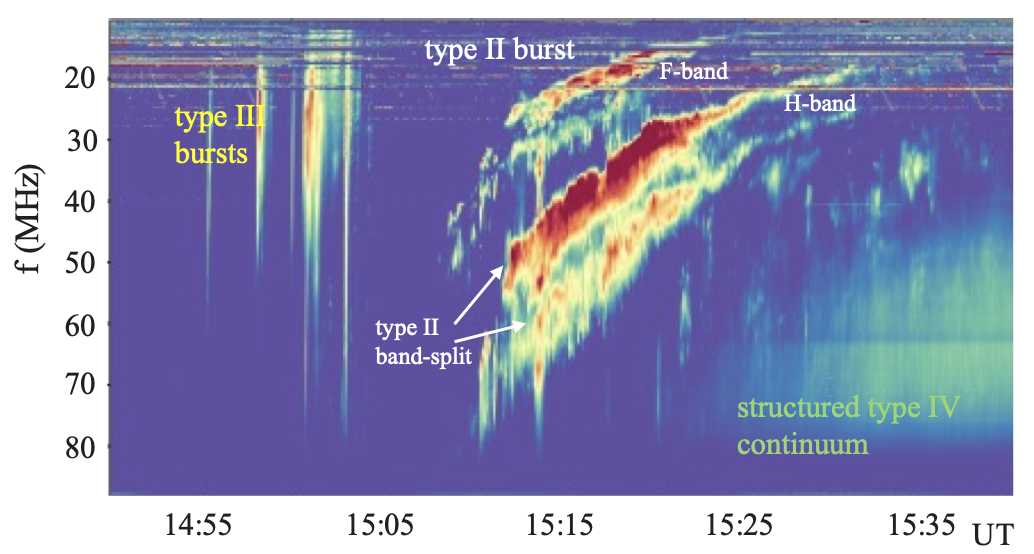}
        \caption{Dynamic spectrum shows typical radio bursts observed during an eruptive event, i.e. type II and type III bursts and continuum type IV radio emission. In addition to different types of bursts, the fundamental and harmonic band (F- and H-band) of type II bursts are marked, together with the two band-split lanes of type II. This radio event was associated with the CME/flare event on August 25, 2014. Adapted from \cite{magdalenic20}.}
    \label{fig:figure10.2}
\end{figure}

\paragraph{Type II Bursts:}

{Type II bursts are slowly drifting emission lanes observed at metric to decametric wavelengths in dynamic spectra at the fundamental and/or harmonic of the plasma frequency \citep{ma96,ne85, Kumari2023b}. The fundamental and harmonic emission bands often show a split, so-called band-split \citep[e.g.][]{Smerd1975, vr01,vr02, Magdalenic2002, kumari2017b, kumari2019} and numerous fine structures \citep[e.g.][see Figure \ref{fig:figure10}]{magdalenic20}. For example,`herringbones' are type II fine structures stemming from a type II `backbone', characterized by their fast drift towards either low or high frequencies \citep{holman1983,ca87,ca89, magdalenic02}. Herringbones sometimes also occur without a type II backbone \citep[][]{holman1983,mo19a}. A broad variety of fine structures, with different morphologies and drifts, comprising the emission lanes of type II bursts, was recently reported by \cite{magdalenic20} and \cite{zhang2024}.}

{Type II bursts are generally considered as the radio signatures of propagating shock waves in the solar corona, with the majority of them being CME-driven \citep[e.g.][]{sm70, ma96,ca13, kumari2017a, zu18,  mo19a,2021AdSpR..68.3464U, 2025A&A...697L...9K} and only occasionally of different origin, like e.g. flare generated \citep[see e.g.][]{Magdalenic2008, Vasanth2024}. Herringbones, fine structures of type IIs represent signatures of individual electron beams escaping the shock, sometimes in opposite directions \citep{zl93, ca13, mo19a}. Unlike the backbone of type II bursts, herringbones have only been imaged on rare occasions \citep[][]{ca13, mo19a, morosan2022} and were found to closely follow the propagation of extreme-ultraviolet (EUV) waves in the low corona and fast lateral expansion of the CME at higher altitudes. The EUV wave represents a fast-mode wave or shock wave that propagates in the low solar corona ahead of the expanding CME  \citep{long2008, kienreich2009, warmuth2015, 2021SoPh..296...62M}. \citet{morosan2022} suggested that the close association of herringbones with the EUV wave and CME expansion points to the formation of a large-amplitude wave that undergoes steepening into a shock during the onset of the CME eruption \citep[][]{Vrsnak2000a, Vrsnak2000b}.}

Several recent studies have focused on determining the origin of band splittings in coronal type II radio bursts, utilising novel high-spatial resolution observations. As mentioned earlier, band-split is a characteristic of type II bursts where emission lanes consist of two or more bands of similar drift rates and morphology \citep[e.g.][]{vr01, magdalenic20}. Recently available high-resolution radio imaging observations in the metric range revealed that these two bands originate in neighboring but distinct regions upstream of the shock \citep[e.g.][]{bhunia2023, zhang2024b,normo2025}. This is contrary to the scenario proposed by \citet{Smerd1975}, suggesting that these bands would be emitted simultaneously upstream and downstream of the shock. On the other hand, the study of interplanetary type II radio bursts \citep[e.g.][]{vr01} clearly shows that, at least in the case of some type II bursts, the band-split is most probably emission from the upstream and downstream shock regions, in accordance with hypothesis by \citet{Smerd1975}. These results indicate that the split of the type II emission bands might not necessarily always have the same origin, in particularly when different plasma regimes such as low corona and interplanetary space, are considered.

\paragraph{Type III Bursts:}

Type III radio bursts are the most frequently observed type of radio burst in dynamic spectra. They appear as fast-drifting structures often spanning the full frequency range, all the way from the gigahertz to kilohertz range. They are signatures of fast electron beams traveling along open or quasi-open magnetic field lines. The type III generating energetic electrons ($\sim \mbox{5\,--\,50\,keV}$) propagate with speeds of fraction of the speed of light often reaching as large speeds as $\sim$0.3~c \citep{reid2014}. The associated type III radio emission is generated through the plasma emission mechanism \citep{Zheleznyakov1968, Melrose1974}. In this nonlinear process, one of the very frequently appearing instability, so called a bump-on-tail instability generates Langmuir waves which subsequently undergo coalescence and/or scattering, producing radio emission at the fundamental plasma frequency or its harmonics \citep[e.g.,][]{Robinson1978}.

Type III radio bursts observed at GHz to few hundred of MHz range are frequently associated with X-ray emission sources \citep{Kane1972, Arzner2005, Reid17, James23}. Studying both types of emissions provides information about the acceleration sites of the associated electron beams \cite[see e.g.][]{Reid11, Bhunia25}. The type III electron beams often, but not always, propagate into the outer corona along open magnetic field lines. Electron beams that do not escape into the outer corona and instead propagate along closed loop structures are often observed as U- or J-bursts \citep{Reid17a}. Such bursts are produced when coronal loops are dense and extended enough to support the instability required for Langmuir wave generation \citep{reid2014}. 

Metric type III radio bursts have been studied for many decades. Dynamic spectra show frequency drift rates of type IIIs to be typically of about 10~MHz/s at 60~MHz, and scale approximately as $df/dt=0.067 f^{1.23} \rm[MHz]$ \citep{zhang2018AAstatisticType3}. The type III drift rates can serve as a proxy for the estimation of the speed of the associated electron beams. By applying coronal electron density models, one can relate the observed emission frequency to height in the corona and thus trace the structure of the coronal electron density \citep[see e.g.,][]{Dabrowski21}. Imaging of type III bursts helps trace the propagation path of the emission and, indirectly, the electron beams themselves. Radio sources at each frequency correspond to different heights in the corona, enabling the mapping of coronal plasma parameters. The source sizes of type III bursts often, but not necessarily always, increases with decreasing frequency, reflecting beam broadening and propagation effects \citep{Steinberg1985, Saint-Hilaire13}. Recent study of interplanetary type III bursts, by \cite{deshpande25} show rather steady trend of the type III source sizes than the clear increase. 

Type III radio bursts are among the most intense structures of all SRBs. Their intensity is, similarly to type II radio bursts, observed to increase up to approximately 1~MHz and then decrease towards lower frequencies. Type III bursts typically continue down to very low frequencies, around 0.1~MHz, reflecting the propagation of the electron beams into the interplanetary medium. However, not all bursts extend that far. Type III bursts ending frequency depends on the radial expansion of the guiding flux tube, the initial conditions of the electron beam, and strong density fluctuations that reduce wave growth \citep{Reid15}.
Type III bursts propagate from high-density to low-density regions through an inhomogeneous ambient plasma, and as a consequence they often exhibit fine frequency structures in the dynamic spectra, typically appearing as horizontal striae with a fractional bandwidth of $\sim$0.1 \citep[see e.g.][]{2021NatAs...5..796R,Jebaraj22}. Fine structures have been studied for a long time in the metric range \citep{ca87}, and recent work has also reported similar structures at very low frequencies \citep{Jebaraj22, 2025ApJ...985L..27K}. These striations are believed to arise from strong density turbulence in the background plasma \citep{2016ApJ...831..154M, 2017SoPh..292..155M}. Observations show that the striae are most prominent in the fundamental component, indicating that it is more sensitive to density fluctuations \citep{Kontar2017, 2018SoPh..293..115S, 2021NatAs...5..796R, Zhang20}. Fundamental and harmonic components of type III bursts are difficult to observe simultaneously, in particular in the metric range, consequently clear examples of the both bursts component are not very often reported \citep{Wild1959, Mann2018c, 2019ApJ...885...78M, Zhang20, Vocks2023}. Recent high frequency/time resolution observations from PSP have shown that both components can indeed be detected in the interplanetary range, with the fundamental component being dominant most of the time \citep{jebaraj2023, Chen24}. 

\paragraph{Type IV Bursts:}
Type IV bursts are long-duration, broadband continuum emissions associated most commonly with CMEs \citep{Webb2012, Robinson1986, Kumari2021, 2022SoPh..297...98K, 2024ApJ...971...86M}.
These emissions can be observed over a wide range of wavelengths spanning from decimetric to metric wavelengths \citep{Pick1986}. 
Based on their appearance in the dynamic spectrum, they are divided into two main sub-categories -- moving type IV bursts (IVm) and the stationary type IV bursts (IVs) \citep{bo57}. Although historically number of different subcategories of the type IV continuum were presented by Kundu, last few decades mostly two main sub-categories are discussed.
While the envelope of the IVm bursts show frequency drift, indicating an outward motion of the associated radio source from the Sun, the IVs bursts do not show any drift, indicating a stationary radio source \citep{Pick1986}.
Schematic examples of IV bursts are shown in Figure \ref{fig:figure10}. These bursts can originate from either stationary or moving radio sources \citep[for a review, see][]{1998ARA&A..36..131B}.

The source of the stationary continuum, so called IVs burst remains rather static in the corona above the active region or in the leg of the post-flare loop systems, and are typically interpreted as emissions from the non-thermal electrons present in association with a flare and the lift-off of a CME \cite[see e.g.][]{salas2020polarisation}. Consequently, type IVs continuum is often considered as a key element to understand the reconstruction in the solar corona after the flare/CME events. Studies of the type IV sources show that the basic source structure of IVs bursts is columnar, i.e., the sources at different frequencies align, at the leg(s) of the erupting flux rope \cite[see e.g.][]{Kai1985, salas2020polarisation}. High brightness temperature (typically $\ge 10^9 K$) and strong left-handed circular polarization (near $100\%$) suggest the IVs to originate through the plasma emission mechanism \citep{Melrose1975}. However, recently some studies report that the emission mechanism of long-duration IVs continuum is an electron cyclotron maser (ECM) process \citep{Liu2018}.


\citet{bastian2001} were the first to report the existence of a 'radio CME', observed as an ensemble of loop structures imaged by the Nan{\c c}ay Radioheliograph. The loops extended to distances of up to 3~R$_\odot$ at 164~MHz and were made visible at radio wavelengths by synchrotron-emitting electrons. Radio imaging and spectroscopy indicate that these moving sources can mark confined populations of accelerated electrons behind the CME or inside CME cores and erupting filaments \citep[e.g.][]{mo19b, vasanth2019, morosan2021, klein2024} and in rare cases, these electrons can be trapped inside the CME\citep{bastian2001, chen2025}. Some IVm bursts can also be composed of fine structures, i.e., narrow-band, highly polarized and/or rapidly varying emission, \citep{Gary1985,vasanth2019,Morosan2020b}
considered to be generated by the plasma emission mechanism. 


\begin{table}[ht]
\centering
\footnotesize
\setlength{\tabcolsep}{4pt}
\renewcommand{\arraystretch}{1.3}

\caption{Summary of the types of solar radio bursts.}
\begin{tabular}{p{2cm} p{3.5cm} p{2.5cm} p{2.5cm} p{3.5cm}}
\hline
\textbf{Type} & 
\textbf{Characteristics} & 
\textbf{Duration} &
\textbf{Frequency Range} &
\textbf{Associated Phenomena} \\
\hline

\textbf{I} &
Short, narrow-band bursts occur in large numbers, usually with an underlying continuum. &
Single burst: $\sim$1 s; Storm: hours–days &
80–200 MHz &
Active regions, flares, eruptive prominences \\

\textbf{II} &
Slow-drift bursts, often with strong second harmonic. &
3–30 min &
Fundamental: 20–150 MHz &
Flares, proton emission, MHD shock waves \\

\textbf{III} &
Fast-drift bursts occur singly, in groups, or in storms (with or without continuum). Often with a second harmonic. &
Single: 1–3 s; Group: 1–5 min; Storm: minutes–hours &
10 kHz–1 GHz &
Active regions, flares \\

\textbf{IV} (Stationary) &
Broadband continuum with fine structure. &
Hours–days &
20 MHz–2 GHz &
Flares, proton emission \\

\textbf{IV} (Moving) &
Broadband continuum with slow frequency drift. &
30 min–2 h &
20–400 MHz &
Eruptive prominences, shocks \\

\textbf{V} &
Smooth, short-lived continuum; follows type III; never occurs alone. &
1–3 min &
10–200 MHz &
Same as Type III bursts \\
\hline
\end{tabular}
\label{tab:table3}
\end{table}

\paragraph{Type V Bursts:}
The most rarely observed radio bursts are type V solar radio bursts. This broadband continuum emission occurs at frequencies generally lower than 200~MHz. It typically appears immediately following a group of Type III bursts and lasts from about one to a few minutes, most often at metric–decametric frequencies (commonly below $\sim$120 MHz). About 45$\%$ of the type V events at 25 MHz accompany type III groups \citep{Daene1966}, though this fraction likely reflects instrumental sensitivity. Combined type III/V activity correlates more strongly with hard X-ray flares than single type III bursts, indicating that they may trace more energetic particle acceleration \citep{Stewart1978}. Some cases show a temporal gap between the two type IIIs and V, forming detached events.

Whether Type V bursts are a genuinely separate class or a manifestation of blended, unresolved type III activity remains contentious \citep{morosan2022b}. Imaging data show that Type V sources occur at comparable coronal heights to Type III sources and show similar displacement in source position with frequency to that of the type IIIs, but their emission regions are often displaced by several tenths of a solar radius. They also decay more slowly and radiate less directionally. Associations between U-shaped type III trajectories and subsequent type V continua indicate that closed magnetic structures can play a role in the generation of these bursts. Their circular polarization is generally weak ($\geq 10\%$) and often opposite in sense to the preceding type III emission, especially when the type III and V sources are spatially well separated \citep{Dulk1980}.

The emission mechanism of type V bursts remains unresolved. Though the emission mechanism was proposed to be gyrosynchrotron of the type III emitting electron beam \citep{Wild1959}, it was rejected early due to spectral inconsistencies. Plasma emission from electrons redistributed or confined in coronal loops has been widely discussed \citep{Weiss1965, Zheleznyakov1968}. However, beam evolution models, ranging from gap-type distributions \citep{Melrose1975} to upper-hybrid wave coalescence and ECM-related scenarios \citep{Winglee1986}, have also been proposed. Several instability-driven scattering mechanisms, including whistler scattering \citep{Melrose1974}, ion-cyclotron–driven isotropization \citep{Benz1974}, and the electron firehose instability (EFI) \citep{Paesold1999,Paesold2003}, may produce the anisotropies required for coherent emission. Recent in situ detection of EFI-like signatures \citep{Cozzani2023} lends support to these ideas. Overall, Type V bursts appear to involve electron populations that evolve differently from those producing narrow beams, which are associated with Type III bursts; however, high-resolution observations are still needed to establish their exact nature.


Figure~\ref{fig:figure10} shows the different types of solar radio bursts and their characteristic signatures in time–frequency space. Table~\ref{tab:table3} summarises the different types of solar radio bursts, their characteristic signatures, and the physical emission mechanisms associated with each class.

\subsection{Probing Coronal Plasma with Solar Radio Bursts}

Solar radio bursts provide a powerful remote probe of the electron beams and
plasma conditions in the solar corona: their brightness, apparent size, and
spectral properties encode how energetic electrons propagate, interact with
turbulent plasma, and convert their energy into coherent radio emission \citep{1985ARA&A..23..169D}.

\begin{figure}
    \centering
    \includegraphics[width=0.8\linewidth]{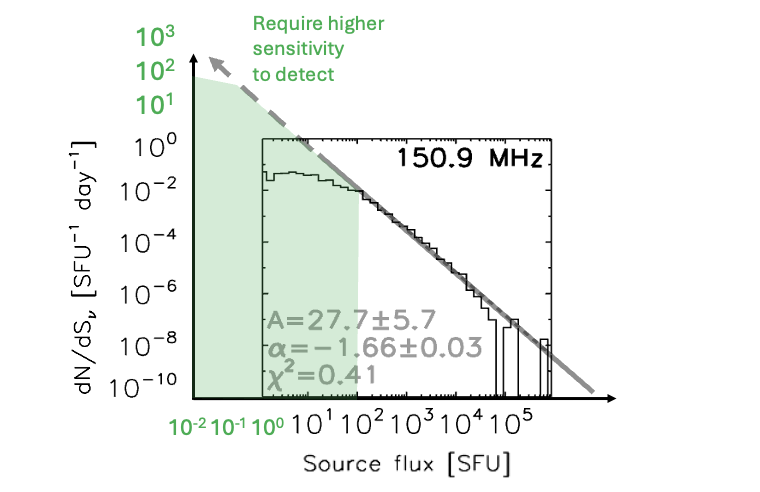}
    \caption{Differential distribution of solar radio burst flux densities at 150.9~MHz \citep{saint2012decade}. The solid line indicates a power-law fit to the observed events, while the shaded region indicates the large, currently unobserved
    population of weak bursts that would require higher sensitivity instruments to be clearly detected.}
    \label{fig:burst_flux_distribution}
\end{figure}

\subsubsection{Brightness}
A central question in the coronal heating problem is whether the energy released by numerous small radio bursts can rival, or strongly contribute to, that of the rarer, very bright events. This can be addressed statistically by measuring the distribution of burst fluxes.  Figure~\ref{fig:burst_flux_distribution} shows the differential number distribution of solar radio burst flux densities $dN/dS_{\nu}$ at 150.9~MHz.  The events, observed over several decades, follow a power-law in flux, with a best-fit slope $\alpha \simeq -1.66$.  The shaded region indicates the regime where the burst fluxes fall below the current detection threshold.  In this range, the expected occurrence rate of radio bursts steeply rises, but the individual bursts remain unresolved in the dynamic spectra. If the power-law continues into this low-flux domain, the cumulative energy carried by the undetected population may be comparable to (or even exceed) that of the bright, easily observed bursts. Improving the instrumental sensitivity and dynamic range is therefore essential for quantifying the true burst energy budget and assessing its contribution to coronal heating.

 \begin{figure}
 \includegraphics[width=0.99\textwidth]{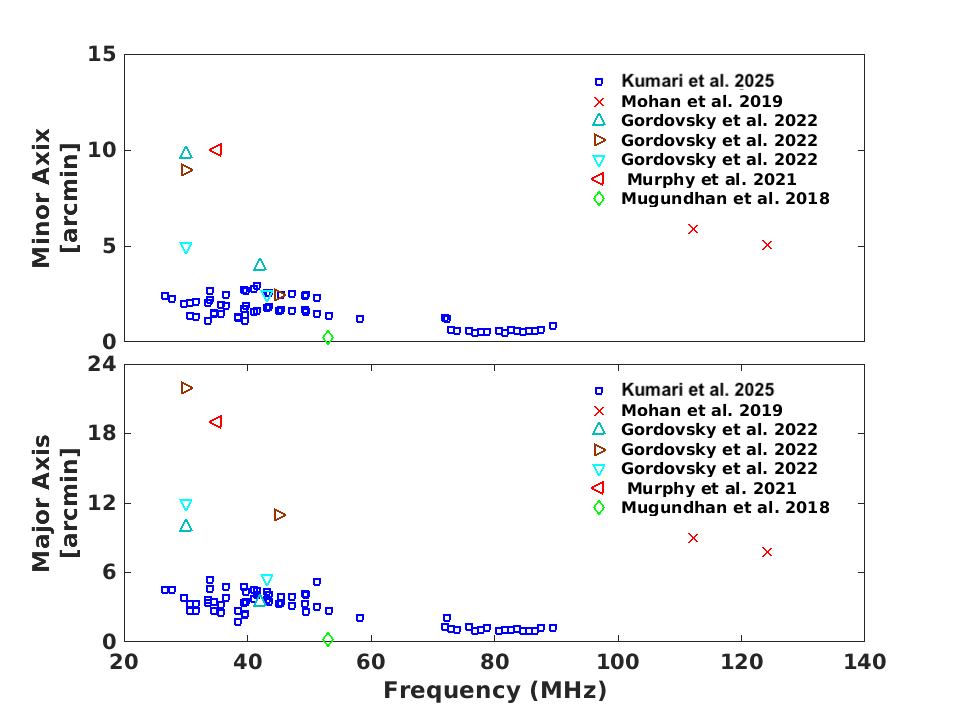}
 \caption{Combined plot of radio burst source sizes from a number of previous studies. 'Blue' squares present the source sizes for fundamental-harmonic lanes of three type II radio bursts. Notably, the source sizes (both minor and major) obtained in the present study are smaller than those in all previous studies using different instruments and observation modes (except \cite{Mugundhan2018}). The figure has been adopted from \cite{kumari2025}.}
  \label{Fig:source_sizes}
\end{figure}

\subsubsection{Radio Bursts Source size}
The apparent size of a radio burst encodes crucial information about the characteristics of the coronal plasma.  At frequencies < 300 MHz, radio waves are notably affected by scattering due to the density fluctuations along the line of sight, which then can broaden and distort the intrinsic source size and structure. 
High-resolution imaging observations over a wide range of frequencies enable us to measure how the source size and morphology evolve with height in the corona, allowing us to separate intrinsic source properties from propagation effects  \citep{mondal2025observation}.  Comparison of observations and forward models of radio-wave scattering and refraction can, to a certain extent, constrain the amplitude and spatial spectrum of coronal density fluctuations, as well as the physical size of the emitting region.

Using LOFAR's large baselines, notable advances have been made in quantifying the source sizes of type II radio bursts.  Figure \ref{Fig:source_sizes} from \cite{kumari2025} presents a combined plot of the sizes of radio bursts observed in various studies to date. These studies include type II bursts observed with both LOFAR's interferometric mode with a maximum baseline of 80~km \citep{kumari2025} and LOFAR's tied-array beam observations of type II radio bursts with a baseline of $\sim$2~km \citep{2022ApJ...925..140G}. Other radio burst source sizes are also included, such as type III bursts observed with MWA with a baseline of  $\sim$3~km \citep{Mohan2019}, type III radio bursts observed with LOFAR's interferometric mode \citep{2021A&A...645A..11M} and VLBI (200~km) of a noise storm \citep{Mugundhan2018}. The source sizes obtained by \citep[][]{kumari2025} are smaller than those in previous studies except that of \cite{Mugundhan2018}, where the baseline was the largest ($\approx 200$~km). Another study by \citet[][]{morosan2025} demonstrated that increasing the interferometric baseline for snapshot imaging with LOFAR also decreases the size of the observed sources, allowing for more details to be resolved in the images. Additionally, an elongated radio source was resolved into two separate components when using LOFAR's longer baselines. These spatial resolution observations also allowed for the determination of rippling scales at the shock front of the order of $\sim 10^5$~km \citep{morosan2025b}. 

\begin{figure}
    \centering
    \includegraphics[width=0.6\linewidth]{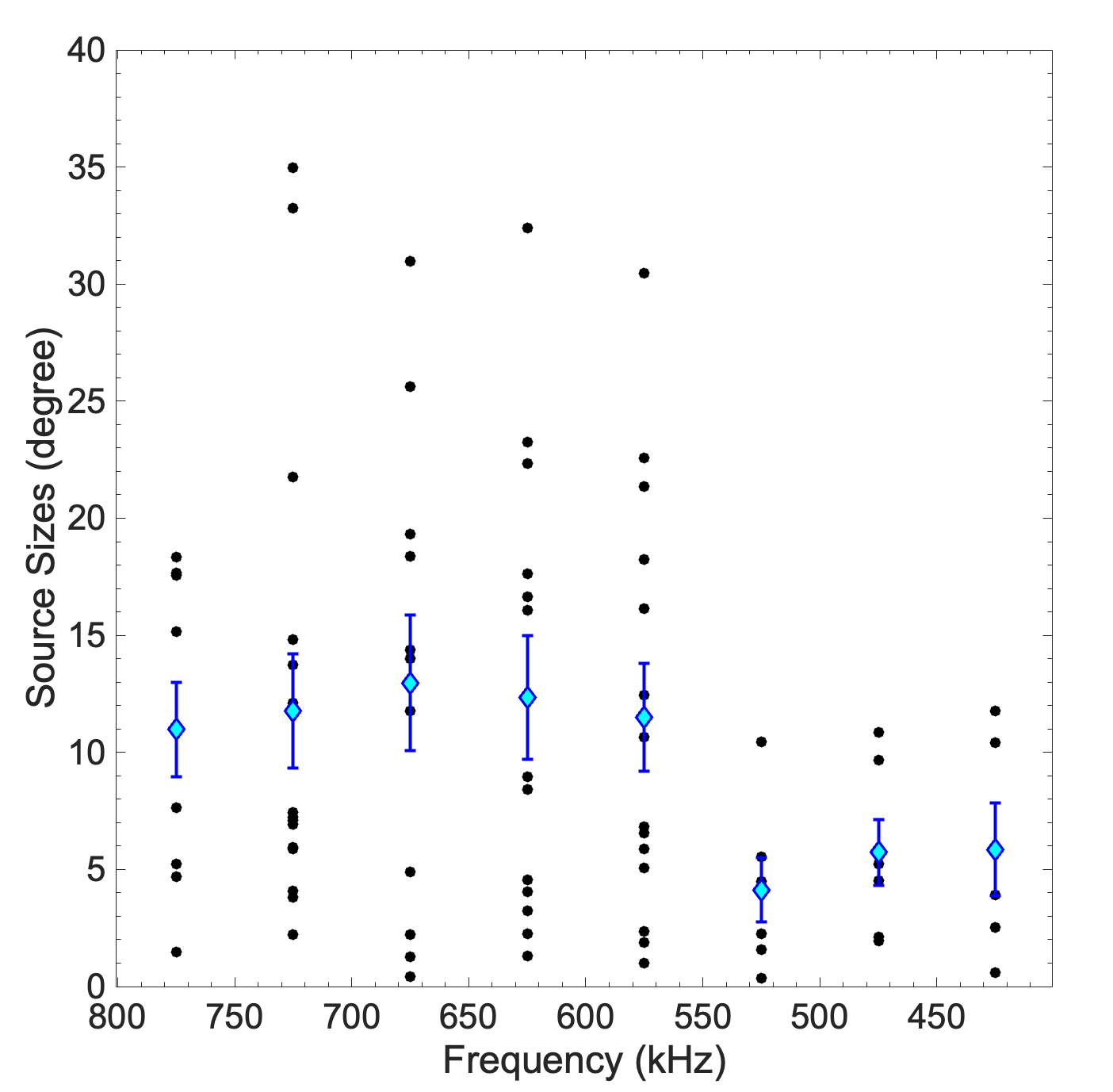}
    \caption{ Variation of radio source sizes with frequency for interplanetary type III radio bursts from the recent study by \cite{deshpande25}. The blue error bars map the standard deviation estimated from bootstrapping and the cyan colored points represent the mean calculated using the bootstrapping method. Adapted from \cite{deshpande25}.}
    \label{fig:figure11a}
\end{figure}

The recent study of interplanetary type III radio burst sources by \cite{deshpande25} employs direction-finding observations and the radio triangulation method in estimating the radio source sizes. The authors demonstrate that type III source sizes are relatively constant (Fig.~\ref{fig:figure11a}) in the frequency range of 800~MHz to 400~MHz. This suggests that scattering effects are likely not dominant in this plasma regime. On the other hand, the authors also estimated level of the density fluctuations using independently the in situ observations by PSP, and found that the level of the density fluctuations is rather low, of the order of 5~-~6~ $\%$  which is in accordance with the rather stable type III radio source sizes obtained with the radio triangulation method.

\begin{figure}
    \centering
    \includegraphics[width=1\linewidth]{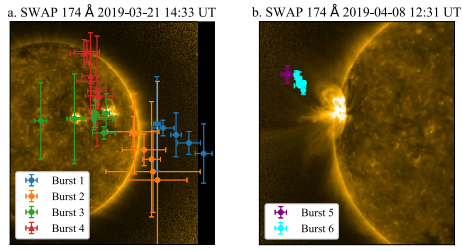}
    \caption{Centroid positions of solar radio bursts derived from tied–array imaging with a low–frequency radio interferometer, overlaid on an EUV image of the Sun. The centroids obtained at multiple frequencies reveal the spatial evolution of burst source regions through the corona. \citep{2022SoPh..297...47M}.}
    \label{fig:figure12}
\end{figure}

\begin{table}[ht]
\centering
\footnotesize
\setlength{\tabcolsep}{4pt}
\renewcommand{\arraystretch}{1.3}

\caption{Summary of the physical diagnostics provided by solar radio burst observations and their complementarity with measurements at other wavelengths. }
\begin{tabular}{p{2.5cm} p{4.5cm} p{7cm}}
\hline
\textbf{Observation from Radio Bursts} & 
\textbf{Physical Information Provided} &
\textbf{Complementarity with Other Wavelengths} \\
\hline

\textbf{Plasma Emission (Types I--V)} &
Electron density, plasma frequency, height of emission, coronal structure; diagnostics of shock speeds (Type II) and electron beam speeds (Type III). &
Complements EUV/X-ray by probing regions where EUV is optically thick or faint; reveals dynamics beyond coronagraph field-of-view; provides density diagnostics not accessible in EUV imaging. \\

\textbf{Electron Acceleration Signatures} &
Fast electron beams (Type III), trapped electrons (Type IV), microbursts, turbulence. &
Hard X-rays observe thick-target bremsstrahlung from footpoints; radio observations, on the other hand, reveal coronal acceleration and transport, providing a coronal counterpart to HXR sources. \\

\textbf{Shock Waves and CME Diagnostics} &
CME-driven shock speed, Mach number, shock geometry, height–time profile (Type II bursts). &
Complements coronagraph CME observations by detecting shocks even when CME fronts are faint; links coronagraph kinematics to shock formation height and SEP production. \\

\textbf{Magnetic Field Diagnostics} &
Gyrosynchrotron emission (GHz), polarization, magnetic field strength and topology, coronal B-field evolution. &
Optical/EUV magnetography only measures photospheric fields; radio provides direct measurements of coronal magnetic fields, filling the gap between photospheric (optical) and in-situ measurements. \\

\textbf{Thermal and Non-thermal Energy Release} &
Flare continua, gyrosynchrotron spectra, accelerated electron distributions, energy partition. &
Soft X-rays measure thermal plasma; radio provides non-thermal signatures, allowing a complete energy budget of flares when combined with GOES/SXR. \\

\textbf{Large-scale Coronal Propagation} &
Tracing electron beams and shocks through the corona and heliosphere (decametric–kilometric bursts). &
Complements in-situ spacecraft (Wind, PSP, Solar Orbiter) by linking remote-sensed radio bursts with in-situ particle and wave measurements along magnetic field lines. \\

\textbf{Heliospheric Imaging (Spacecraft)} &
Low-frequency radio triangulation, direction-finding, and source location of Type II/III bursts in the heliosphere. &
No other wavelength can remotely detect electron beams and shocks beyond ~20 $R_\odot$; it complements white-light heliospheric imagers by probing invisible shocks and magnetic connectivity. \\
\hline
\end{tabular}
\label{tab:table4}
\end{table}

\subsubsection{Emission mechanism}
\begin{figure}
    \centering
    \includegraphics[width=\linewidth]{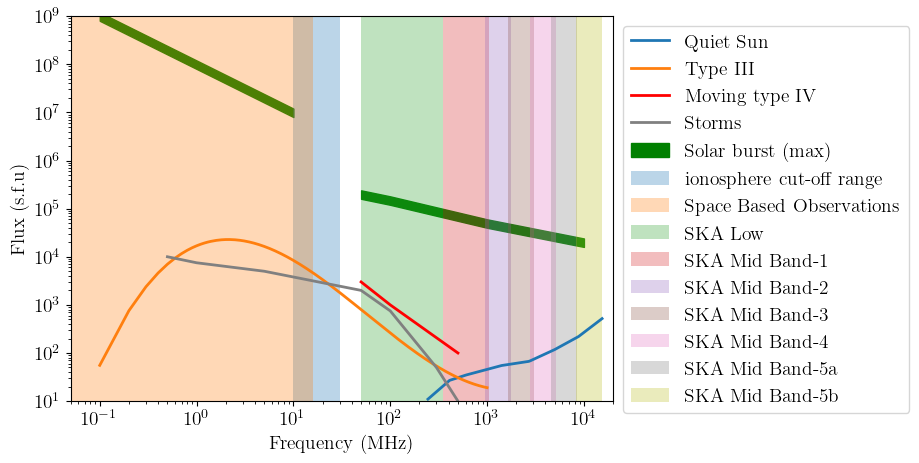}
    \caption{The variation of flux density with frequency for different types of solar radio bursts. This illustrates how burst intensity changes across the observed spectral range.}
    \label{fig:figure13}
\end{figure}

The radio emission mechanism links the observed properties of the bursts to the underlying energetic particles and plasma conditions.  Most metric and decametric solar radio bursts are produced by coherent plasma emission. The nonthermal electron beams or unstable electron distributions drive Langmuir waves, which are converted into escaping electromagnetic radiation at the local plasma frequency and/or its harmonics. The resulting spectra is of a narrow bandwidth, very high brightness temperatures, and characteristic frequency–time drift rates that reflect the electron propagation through the coronal plasma of a changing density. By combining information on the spectral characteristics of the bursts such as e.g. drifts, band split, fine structures characteristics and polarization level, with the constraints on source position and size obtained from imaging observations, we can distinguish between different
emission scenarios. That allows us to infer key parameters of the electron population and the ambient corona, such as density, magnetic field strength, and plasma turbulence level. For a detailed discussion of the emission mechanisms, see the chapter \cite{Patra01.2026.SKA}  in this book. Table~\ref{tab:table4} provides a summary of the diagnostics that are possible employing solar radio bursts. We also highlight the complementarity of radio observations with observations across the other wavelength domains. Figure~\ref{fig:figure12} shows a combined plot of the radio observations, AIA EUV images, and white–light coronagraph data. This highlights the complementarity of radio observations with measurements across other wavelength domains. The figure illustrates the temporal and spatial relationship between radio emission and coronal structures observed in EUV. The radio signatures trace energetic electron populations and shock-related processes, while the EUV images reveal the evolving morphology of the coronal plasma. The combined view demonstrates how radio diagnostics provide direct insight into particle acceleration and plasma emission processes that are not accessible through thermal imaging alone.
Figure~\ref{fig:figure13} shows the variation of flux density with frequency for different classes of solar radio bursts across the SKA frequency bands. The figure emphasizes how distinct burst types dominate different frequency regimes, reflecting changes in emission mechanism and source height in the solar atmosphere. Coherent plasma emission processes dominate at low frequencies, producing high brightness temperatures, while incoherent gyrosynchrotron emission becomes increasingly important at higher frequencies. This frequency-dependent behavior underscores the importance of broad spectral coverage to fully characterise solar eruptive events.

\subsection{Current Observations and Challenges}

While many solar observing instruments are present worldwide, there is a dearth of dedicated solar imaging instruments, both at low and high frequencies. There is also a 15-20 UT time gap where neither broadband dynamic spectrometers nor imaging telescopes are available. This gap needs to be filled in order to obtain synoptic monitoring of the Sun. Furthermore, instruments that perform imaging are restricted to Stokes I due to difficulties in polarization calibration, as it depends on the telescope's systematic response, the position of the Sun and calibrators in the sky, and the ionosphere at low metric to decametric range frequencies. With advances in ionospheric monitoring, learned from LOFAR, and in polarization calibration, learnt from MWA, some of these challenges can be resolved in the upcoming decade using the SKA.

\section{SKA Capabilities for Solar Science}
\subsection{Solar Radio Monitoring}
SKA, with its unprecedented frequency coverage, will enable monitoring solar bursts from the low corona to the inner heliosphere. SKA Low has 256 dual-polarized log-periodic dipoles operating between 50 and 350 MHz. A single station will have a field of view of $\approx$ 90$^{o}$, enabling solar monitoring of $\sim$ 6 hours. The SEFD of a single SKA station has been measured to be $\approx$ 5.4 kJy at 90 MHz, while it is 3.7 kJy at 310 MHz. Either way, the sensitivity will be at a sub-SFU level, facilitating the detection of weak bursts. Each station can form 48 beams, out of which one Solar beam will allow us to achieve most of the monitoring requirements. 

The data from the monitor beam can be searched for events using the native SKA Low time and frequency resolution. Data containing interesting events can be stored for detailed analysis, while an averaged version of the dynamic spectra can be used for long-term archival purposes.

While obtaining a solar dedicated beam may not be possible with SKA-mid, opportunistic observations can be made using triggers from the SKA-Low monitor beam or from other solar telescope networks after the detection of solar flares. This will be particularly useful for probing impulsive flare-related energy releases in the microwave frequency range.  Table~\ref{tab:table5} summarises the key parameters of SKA1-LOW and SKA1-MID that are most relevant for solar radio physics.

\begin{table}[ht]
\centering
\small
\caption{Key parameters of SKA1-LOW and SKA1-MID relevant for solar radio physics.}
\begin{tabular}{lcc}
\hline
\textbf{Parameter} & \textbf{SKA1-LOW} & \textbf{SKA1-MID} \\
\hline
Frequency range & 50--350 MHz & 0.35--15.3 GHz \\
Primary science & Plasma emission & Gyrosynchrotron, free-free \\
Antenna elements & 131,072 dipoles & 197 dishes (15 m) \\
Stations / Dishes & 512 stations & 133 (SA) + 64 MeerKAT \\
Max. baseline & $\sim$65 km & $\sim$150 km \\
Angular resolution & $\sim$10'' @ 100 MHz & $\sim$0.5'' @ 10 GHz \\
Time resolution & $<$1 ms & $<$1 ms \\
Spectral channels & $>10^4$ & $>10^4$ \\
Instantaneous BW & $\sim$300 MHz & $\sim$5 GHz \\
Polarization & Full Stokes & Full Stokes \\
Solar dedicated & No (but capable) & No (but capable) \\
Key diagnostics & Type II/III, CME shocks & Flares, coronal heating \\
\hline
\end{tabular}
\label{tab:table5}
\end{table}

\subsection{High time and frequency resolution observations}
While solar bursts have been classified into five well-known and understood types (refer to Figure \ref{fig:figure10}), there exists a plethora of fine structure bursts whose association with energy releases in the Sun is not very well understood. The fine structures of the type II and type III bursts can be used to understand the propagation effects of the electron beams and shocks in the solar corona, if they are well resolved. With SKA's frequency resolution of $\approx$ 5 kHz and high time resolution $\sim$ few ms, fine structure bursts can be studied in an unprecedent way. For instance, striae in type III bursts were $\sim$ few 10s of kHz, and lasted a few 10s of ms in duration. These bursts can be sampled to high resolution, and various correlation analyses can be performed on them to understand the nature of plasma turbulence. Similarly, fine structure in type II bursts can also be used for remote sensing of the plasma around the shock front, giving rise to the radio emission. The fine structures often superimposed on the type IV continuum, such as fibers, spikes, pulsations, or zebra patterns \citep{Slottje1972, Magdalenic2006} when fully resolved in both temporal and frequency domains, allow an estimate of the plasma characteristics and reveal information on the associated small-scale process.  

\subsection{Imaging Spectroscopy and Synthesis Imaging Observations}
Imaging spectroscopy is a crucial technique that has gained increasing use in recent years, thanks to the availability of excellent wide-band antennas and affordable signal processing hardware. LOFAR has pioneered this by forming multiple tied-array beamlets\footnote{Beams at different frequencies are beamlets in LOFAR parlance}, which has allowed the localization of radio-emission regions across varying heliocentric distances. 
Imaging observations of the Sun have been performed using both dedicated and generic radio interferometers. The solar disk in optical wavelengths has a diameter of $\approx$ 32$'$; this becomes larger at low frequencies. 
Dedicated solar radio interferometers used 'T' shaped grating arrays to achieve a `filled' uv plane. This may not be suitable for wide-band arrays due to the chromaticity of the grating response. 

\subsection{Solar Long Baseline Interferometer Observations}

\begin{figure}[ht]
    \centering
    \includegraphics[width=0.95\textwidth]{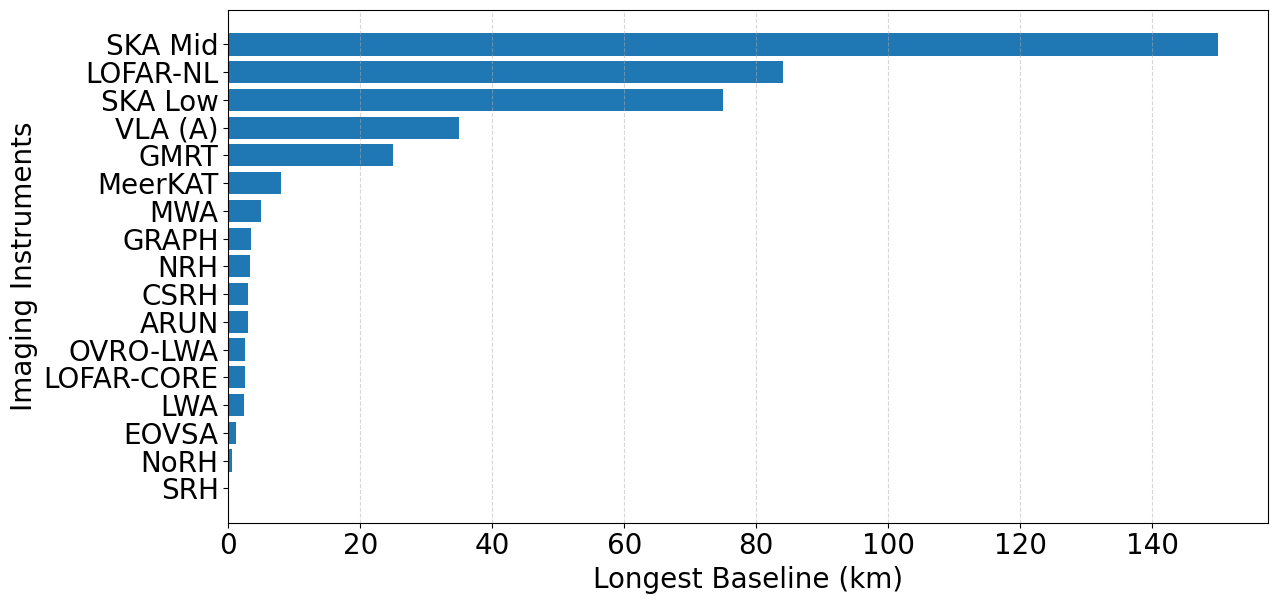}
    \caption{Comparison of the maximum baselines of major radio interferometers used for solar and non-solar radio imaging. Arrays such as LOFAR, VLA, MeerKAT, GMRT and the upcoming SKA provide long-baseline capabilities, enabling high angular resolution essential for studying fine-scale coronal structures and solar radio burst source locations.}
    \label{fig:figure14}
\end{figure}

Solar long-baseline interferometric observations with the Square Kilometre Array (SKA-Low and SKA-Mid) will provide unprecedented capabilities for high-dynamic range, high-fidelity imaging of the solar corona across a wide frequency range (50 MHz to 15 GHz). The availability of baselines extending up to tens of kilometres enables spatial resolutions of a few arcseconds at low radio frequencies and sub-arcsecond imaging in the microwave range, surpassing all existing solar radio interferometers. In comparison, current instruments such as the MWA, LOFAR, and EOVSA offer important but more limited baselines, resulting in coarser angular resolutions and lower imaging dynamic range. While MWA and LOFAR provide excellent low-frequency coverage and snapshot spectroscopic imaging, and EOVSA excels in broadband microwave imaging spectroscopy for flare diagnostics, the SKA will combine these capabilities within a single instrument, thus enabling comprehensive diagnostics of coronal plasma, magnetic fields and particle acceleration over multiple spatial scales. Consequently, the SKA is expected to establish a new benchmark for long-baseline solar radio interferometry and substantially enhance the global capability for solar physics and space weather research. Figure~\ref{fig:figure14} shows the comparison of the longest baselines for major global radio interferometers relevant to solar radio physics and high–resolution imaging.

\subsection{Technical Challenges and Path Forward}

SKA has the advantage of a larger frequency bandwidth for solar radio bursts tracking and increased temporal and spectral resolution. Compared to LOFAR, the SKA is also expected to have increased sensitivity over longer baselines, especially with SKA-Low at low frequencies. SKA-Low, combined with SKA-Mid, can also observe Type II and other radio bursts at much higher frequencies compared to LOFAR. High-frequency type II bursts, starting at frequencies above 500 MHz, have been reported on several occasions \citep[e.g.,][]{kumari2017a, vasanth2025}, indicating the presence of early low coronal shock waves. The larger observing bandwidth of SKA will enable the tracking of electron beams closer to the erupting region and over longer distances, helping to investigate the formation of coronal shocks. The increased spatial resolution of SKA can determine smaller source sizes or resolve multiple bursts contributing to extended emissions, and in turn be used to deduce properties of the shock \citep[e.g.][]{morosan2025b}. There is a detailed discussion regarding the chapter titled 'State-of-the-art Observation, Calibration, and Imaging Framework for Solar and Heliospheric Sciences with SKAO' in this book by \cite{Oberoi01.2026.SKA}.

\section*{Acknowledgments}

A. K., M. V., D. E. M., J. M. and K. D. contributed equally to this work. We acknowledge the CESRA community website curated by Dr.~Eduard Kontar for providing valuable resources related to solar radio observations. A. K. acknowledges support from the ANRF Prime Minister's Early Career Research Grant (PM-ECRG) Program. J. M. acknowledges financial support from the BRAIN.be project SWIM, University start-up grant 3E220031, and the FEDtWIN project PERIHELION. K. D. acknowledges the PhD grant from the Royal Observatory of Belgium and the C1 project UnderRadioSun at KU Leuven. We are grateful to the various observatories and instrument teams whose publicly available datasets and documentation enabled the compilation of instrument characteristics and the generation of comparative plots. We also acknowledge the use of the SunPy Python package, along with the broader open-source scientific Python ecosystem.

\section*{Appendix}

Table~\ref{tab:lightcurve} presents a summary of the major light–curve instruments used for solar radio observations. Additional details of these instruments are provided in Section \ref{sec:section2.1}. Table~\ref{tab:spectrograph} summarises the key solar radio spectrographs and spectropolarimeters currently in operation. Further technical descriptions and references are discussed in Section \ref{sec:section2.2}. Table~\ref{tab:Imaging} provides an overview of the principal radio imaging facilities that contribute to both solar and non-solar science. More detailed information on these telescopes is included in Section \ref{sec:section2.3}. Together, these tables provide a comprehensive snapshot of the global solar radio observational infrastructure, encompassing time–series monitoring, broadband spectral measurements, and high–resolution imaging. They highlight the diversity of instruments, observing schemes, and frequency coverages that complement each other in studying solar radio bursts and related heliophysical processes.

\begin{table*}[ht]
\centering
\begin{threeparttable}
\caption{Summary of instruments and their operating parameters.}
\begin{tabularx}{\textwidth}{l c X c}
\hline
\textbf{Instrument} &
\textbf{Obs. Time (UT)} &
\textbf{Channels (GHz)} &
\textbf{Data Type} \\
\hline

VIRAC RT-32 & 07--14 & [6.3, 9.3] & RCP/LCP \\

RT-2 / RT-3 & 05--13 & [6.15, 9.35] & Total Intensity \\

SSRT & 07--14 &
[4.5, 5.2, 6.0, 6.8, 7.5] &
Total Intensity \\

SRH & 07--14 &
[4.5, 5.2, 6.0, 6.8, 7.5] &
Total Intensity \\

SST & 17--23 &
[212, 405] &
Total Intensity \\

SSRT SP & 07--14 &
[2.34, 2.57, 2.85, 3.20, 3.63, 4.16, 4.82,
5.62, 6.60, 7.82, 8.75, 10.10, 13.20,
15.70, 19.90, 22.93] &
RCP/LCP \\

Metsähovi & 09--16 &
[11.2] &
Total Intensity \\

NoRP & 22--30 &
[1, 2, 3.75, 9.4, 17, 35, 80] &
RCP/LCP \\

RSTN & 00--24 &
[0.245, 0.410, 0.610, 1.415, 2.695,
4.995, 8.800, 15.400] &
Total Intensity \\

GRIP & 02--08 &
[0.0795, 0.0805] &
Stokes I and V \\

Ondrejov RT3 & 07--14 &
[2.9, 3.1] &
Total Intensity \\
\hline
\end{tabularx}
\label{tab:lightcurve}
\end{threeparttable}
\end{table*}

\begin{table*}[ht]
\centering
\begin{threeparttable}
\caption{Summary of major solar radio spectrographs and their operating parameters.}
\begin{tabularx}{\textwidth}{l c c c c}
\hline
\textbf{Instrument} & \textbf{$F_{l}$ (MHz)} & \textbf{$F_{u}$ (MHz)} &
\textbf{Obs. Time (UT)} & \textbf{Type} \\
\hline

URAN-2          & 8    & 33    & 07--14 & Spectropolarimeter \\
OSRA            & 40   & 800   & 09--15 & Spectrograph \\
SSRT            & 4000 & 8000  & 07--14\tnote{a} & Spectropolarimeter \\
GLOSS           & 35   & 435   & 02--09 & Spectrograph \\
GRASP           & 50   & 500   & 02--09 & Spectropolarimeter \\
HSRS            & 275  & 1495  & 08.5--16 & Spectrograph \\
HiRAS           & 25   & 2500  & 22--24 & Spectrograph \\
Learmonth       & 25   & 180   & 22--24 & Spectrograph \\
ARTEMIS-IV      & 20   & 650   & 08--15 & Spectrograph \\
IZMIRAN         & 20   & 260   & 07--14 & Spectrograph \\
Ondrejov        & 800  & 5000  & 08--14 & Spectrograph \\
Culgoora        & 18   & 1800  & 22--24 & Spectrograph \\
ETH 5m STS      & 1415 & 1432  & 09--15 & Spectrograph \\
ETH 5m Bleien   & 1590 & 1650  & 09--15 & Spectrograph \\
ETH LWA Bleien  & 10   & 80    & 09--15 & Spectropolarimeter \\
NDA             & 10   & 100   & 09--15 & Spectrograph \\
Phoenix-3       & 100  & 5000  & 09--15 & Spectrograph \\
\hline
\end{tabularx}

\begin{tablenotes}
\footnotesize
\item[a] SSRT has seasonal operation: typically 23--10 UT in summer.
\end{tablenotes}

\label{tab:spectrograph}
\end{threeparttable}
\end{table*}

\begin{table*}[ht]
\centering
\begin{threeparttable}
\caption{Summary of major radio telescopes used for solar and non-solar radio imaging.}
\begin{tabular}{lccccc}
\hline
\textbf{Telescope} & \textbf{$F_{l}$ (MHz)} & \textbf{$F_{u}$ (MHz)} & \textbf{Obs. time (U.T.)} &\textbf{Type} & \textbf{Solar Dedicated} \\
\hline
GRAPH        & 40      & 120      & 2.5 \,--\, 9 & Interferometer  & Y \\
CSRH         & 400     & 15000    & 1 \,--\, 8 & Interferometer  & Y \\
DART        & 150     & 450    & 1 \,--\, 8 & Interferometer  & Y \\
SunDish      & 18000   & 26000    & 9 \,--\, 14 & Single Dish     & Y \\
SRH          & 4000    & 8000     & 7 \,--\, 14 & Interferometer  & Y \\
Metsähovi    & 37000   & 37000    & 9 \,--\, 14 & Single Dish     & Y \\
NoRH\tnote{*}        & 17000   &  34000 & 22 \,--\, 6    & Interferometer  & Y \\
EOVSA        & 1200    & 18000   & 14 \,--\, 2 & Interferometer  & Y \\
OVRO-LWA     & 10      & 90      & 14 \,--\, 2 & Interferometer  & Y \\
RATAN-600    & 610     & 30000   & 7 \,--\, 14& Interferometer  & Y \\
NRH          & 150     & 450     & 8 \,--\, 15 & Interferometer  & Y \\
ARUN\tnote{\#}       & 1000       & 12000 & 2.5 \,--\, 9 & Interferometer  & Y \\
\hline
uGMRT         & 150     & 1500     & 2.5 \,--\, 9 & Interferometer  & N \\
MWA          & 80      & 300      & 17 \,--\, 23 & Interferometer  & N \\
LOFAR        & 10      & 220      & 8 \,--\, 15 & Interferometer  & N \\
MeerKAT      & 0.58    & 3500     & 7\,--\, 14 & Interferometer  & N \\
LWA          & 10      & 90       & 20 \,--\, 2 & Interferometer  & N \\
VLA          & 1000    & 50000    & 20 \,--\, 2 & Interferometer  & N \\
SKA-Low\tnote{\#}   & 50      & 350    & 17 \,--\, 23 & Interferometer  & N \\
SKA-Mid\tnote{\#}   & 350     & 15000  &  7 \,--\, 14 & Interferometer  & N \\
ALMA         & 35000   & 950000  & 20 \,--\, 2 & Interferometer  & N \\
\hline
\end{tabular}
\label{tab:Imaging}
\begin{tablenotes}
\footnotesize
\item[]$F_{l}$ and $F_{u}$ are lower and upper frequencies of operation of the aforementioned instruments.
\item[*] NoRH is not operational anymore.
\item[\#] Upcoming/upcoming-generation radio instruments (ARUN, SKA-Low, SKA-Mid).
\end{tablenotes}

\end{threeparttable}

\end{table*}

\bibliographystyle{abbrvnat-maxbibnames4}
\bibliography{chapter} 

\end{document}